\documentclass{emulateapj}
\usepackage{graphicx}
\usepackage{multirow}
\usepackage{color}
\usepackage{amssymb}
\usepackage{natbib}
\defcitealias{sch98}{SFD98}

\shorttitle{Metallicity Gradients Observed by SEGUE}
\shortauthors{Cheng et al.}

\begin{document}

\title{Metallicity Gradients in the Milky Way Disk as Observed by the SEGUE Survey}

\author{Judy Y. Cheng}
\affil{Department of Astronomy and Astrophysics, University of California Santa Cruz, Santa Cruz, CA 95064, USA}
\email{jyc@ucolick.org}

\author{Constance M. Rockosi\altaffilmark{1}}
\affil{UCO/Lick Observatory, Department of Astronomy and Astrophysics, University of California, Santa Cruz, CA 95064, USA}

\author{Heather L. Morrison}
\affil{Department of Astronomy, Case Western Reserve University, Cleveland, OH 44106, USA}

\author{Ralph A. Sch{\"o}nrich}
\affil{Max-Planck-Institute f{\"u}r Astrophysik, Karl-Schwarzschild-Str. 1, D-85741 Garching, Germany}

\author{Young Sun Lee}
\affil{Department of Physics and Astronomy and JINA: Joint Institute for Nuclear Astrophysics, Michigan State University, E. Lansing, MI 48824, USA}

\author{Timothy C. Beers}
\affil{Department of Physics and Astronomy and JINA: Joint Institute for Nuclear Astrophysics, Michigan State University, E. Lansing, MI 48824, USA}
\affil{National Optical Astronomy Observatory, Tucson, AZ 85719 USA}

\author{Dmitry Bizyaev}
\affil{Apache Point Observatory, Sunspot, NM, 88349}

\author{Kaike Pan}
\affil{Apache Point Observatory, Sunspot, NM, 88349}

\author{Donald P. Schneider}
\affil{Department of Astronomy \& Astrophysics, Pennsylvania State University, University Park, PA 16802, USA}

\altaffiltext{1}{Packard Fellow}

\begin{abstract}
The observed radial and vertical metallicity distribution of old stars in the Milky Way disk provides a powerful constraint on the chemical enrichment and dynamical history of the disk. We present the radial metallicity gradient, $\Delta{\rm [Fe/H]}/\Delta R$, as a function of height above the plane, $|Z|$, using 7010 main sequence turnoff stars observed by the Sloan Extension for Galactic Understanding and Exploration (SEGUE) survey. The sample consists of mostly old thin and thick disk stars, with a minimal contribution from the stellar halo, in the region $6 < R < 16$ kpc, $0.15 < |Z| < 1.5$ kpc. The data reveal that the radial metallicity gradient becomes flat at heights $|Z|>1$ kpc. The median metallicity at large $|Z|$ is consistent with the metallicities seen in outer disk open clusters, which exhibit a flat radial gradient at [Fe/H] $\sim-0.5$. We note that the outer disk clusters are also located at large $|Z|$; because the flat gradient extends to small $R$ for our sample, there is some ambiguity in whether the observed trends for clusters are due to a change in $R$ or $|Z|$. We therefore stress the importance of considering both the radial and vertical directions when measuring spatial abundance trends in the disk. The flattening of the gradient at high $|Z|$ also has implications on thick disk formation scenarios, which predict different metallicity patterns in the thick disk. A flat gradient, such as we observe, is predicted by a turbulent disk at high redshift, but may also be consistent with radial migration, as long as mixing is strong. We test our analysis methods using a mock catalog based on the model of Sch{\"o}nrich \& Binney, and we estimate our distance errors to be $\sim25\%$. We also show that we can properly correct for selection biases by assigning weights to our targets.
\end{abstract}
\keywords{Galaxy: abundances, Galaxy: disk, Galaxy: evolution, Galaxy: formation}

\section{Introduction}\label{intro}
\subsection{Metallicity Gradients}
The spatial variation in the metallicity distribution of old stars in the Milky Way disk is linked to the formation and evolution of the Galaxy. The metallicity of stars at a particular place in the disk depends on the gas accretion rate, star formation history, and subsequent evolution at that location. For example, in the simplest picture of ``inside-out" disk formation, low angular momentum gas falls to the center of the halo first, forming stars earlier and becoming chemically enriched much faster than the outer disk (e.g., \citealt{lar76,mat89,chi97,pra00}). In this scenario, heavy element abundances decrease as a function of Galactocentric radius, i.e., the metallicity gradient is negative. The presence of radial flows (e.g., \citealt{lac85,goe92,por00,spi11}) and the nature of the early infalling gas (e.g., \citet{cre10}), however, have significant impacts on the chemical evolution of the disk and can generate a gradient that is weaker or even reversed compared to the simplest picture. Thus, observations of the radial metallicity gradient of the disk are crucial to constraining chemical evolution models (e.g., \citealt{chi01,ces07,mag09}).

The radial metallicity gradient of the Milky Way disk has been measured using a number of different tracers, including Cepheids and open clusters, yielding a value between $\sim -0.01$ and $-0.09$ dex kpc$^{-1}$ (e.g., \citealt{cap01,fri02,che03,luc06,lem08,ses08,ped09,luc11}). These tracers represent the composition of the interstellar gas at the time that they were formed, thus the wide variety of tracers studies probe the metallicity gradient at different times. This simple picture, however, can be complicated by processes that change the orbits of stars, such as dynamical heating from perturbations like spiral structure, molecular clouds, or minor mergers, which can make gradients shallower or wash them out completely.

In addition to the steepness of the gradient, non-linear features in the metallicity distribution provide further observational constraints. Some authors, for example, have noted the presence of a discontinuity in the radial metallicity gradient at a Galactocentric radius of $R\sim 10$ kpc, beyond which the metallicity gradient becomes shallower or flat (i.e., slope close to or equal to zero). This feature has been studied extensively using open clusters (e.g., \citealt{twa97,yon05,car07,ses08}) and Cepheids (e.g., \citealt{and02c,yon06,ped09}). 

Possible explanations for this discontinuity include dynamical interactions within the disk (e.g., \citealt{and04}) and merger or accretion events in the outer disk (e.g., \citealt{yon05}). In the former scenario, the presence of the Galactic bar and spiral arms may influence the star formation rate and flow of gas throughout the disk, affecting the amount of chemical enrichment that occurs at different radii. The latter scenario invokes star formation triggered by small accretion events, which is thought to explain the chemical abundances observed for outer disk open clusters (in particular the enhanced abundances of $\alpha$ and $r$-process elements), which differ from the general Galactic population.

In this paper, we present the radial metallicity gradient $\Delta$[Fe/H]/$\Delta R$ of the Milky Way disk, as a function of height above the plane $|Z|$, using a sample of 7010 field stars from the Sloan Extension for Galactic Understanding and Exploration (SEGUE; \citealt{yan09}), part of the Sloan Digital Sky Survey (SDSS; \citealt{yor00}). The sample covers Galactic coordinates $6< R < 16$ kpc, $0.15 < |Z| < 1.5$ kpc, where $R$ is the cylindrical Galactocentric radius and $|Z|$ is absolute distance from the plane. Our field stars are older than the open clusters and Cepheids used in previous gradient measurements in the literature and serve to extend the observations of the metallicity distribution of the disk to older tracers, which can provide constraints on the strength of the gradient at early times and how much radial mixing occurred.

Additionally, we examine whether a discontinuity in the radial metallicity gradient exists in the old disk stars. While the distances derived for individual field stars are less accurate than for other tracers, our sample is sufficiently large to divide into bins of $|Z|$, which allows for an examination of the metallicity distribution in the disk as a function of both $R$ and $|Z|$. The distinction between $R$ and $|Z|$ is important, as many of the outer disk tracers in the literature are also located far from the midplane, and the question of whether the reported trends are a function of $R$, $|Z|$, or both, needs to be assessed.

\subsection{Thick Disk Formation}
This work is further motivated by the idea that the metallicity gradient of the old disk may be used as an observational constraint to distinguish between possible formation mechanisms for the thick disk. Traditionally, the Galactic disk can be thought of as the sum of two components: a thin disk of young metal-rich stars and a thick disk of older, more metal-poor stars (e.g., \citealt{gil95,chi00,ben04b,ive08}). The existence of the Milky Way's thick disk was first noted by \citet{yos82} and \citet{gil83}, and thick disks with similar kinematics, structure, and stellar populations have since been observed to be a common feature in nearby spiral galaxies \citep{dal02,yoa05,yoa06,yoa08b}.

The ubiquity of thick disks in external galaxies, as well as the similarity of their properties, suggests that whatever process is responsible for their existence is important in the formation and evolution of disk galaxies. Furthermore, because the thick disk is old, the properties of its stars can be used as a \textquotedblleft fossil record" of the disk's early formation. How stars end up in a thick disk, far from the plane of the Galactic disk, remains an open question. 

Several mechanisms for thick disk formation have been proposed, four of which are discussed below. Within the context of the hierarchical structure formation predicted by $\Lambda$CDM cosmology, a thick disk may arise through (1) the puffing up or vertical heating of a pre-existing thin disk during a minor merger (e.g., \citealt{vil08,rea08,kaz08,kaz09,pur09,bir11}); (2) the direct accretion of stars formed in satellites that merged with the Galaxy \citep{aba03}; or (3) star formation in an early turbulent disk phase during a period of high gas accretion (e.g., \citealp{bro04,bro05,bou09}). Even in the absence of cosmological accretion, a thick disk may also arise through (4) radial migration of stellar orbits (e.g., \citealt{sch09a,sch09b,loe11}).

Each of these scenarios is motivated by both theory and observations. Halo merger histories in cosmological N-body simulations suggest that the types of mergers required by scenarios 1 and 2 are common; \citet{ste08} estimate that $70\%$ of Milky Way-sized halos have experienced a 1:10 merger within the last 10 Gyr. Streams in the halo of the Milky Way (e.g., \citealp{new02,bel07}) and other galaxies (e.g., \citealp{iba01,mar10}) provide observational evidence of such accretion events. Other recent work, however, has emphasized the importance of smooth gas accretion in the growth of disk galaxies \citep{bro09,dek09}, which can lead to a turbulent disk with high velocity dispersion as in scenario 3. Observational support for this picture include the clump-cluster galaxies \citep{elm05} and the thick chain and spiral galaxies \citep{elm06} seen at high redshift. Lastly, \citet{sel02} and \citet{ros08a,ros08b} showed that resonant interactions between stars and transient spiral waves can change the radii of stellar orbits while keeping them on circular orbits, leading to the kind of radial mixing necessary for scenario 4. Observations of nearby stars indicate that radial migration is an important process that may shape the correlations between the kinematics, metallicities, and ages of stars in the solar neighborhood \citep{hay08}. 

While there is evidence that the mechanisms described above---minor mergers, early gas accretion, and radial migration---are at play in galaxy formation, the question of which mechanism, if any, is the \textit{dominant} force behind thick disk formation remains unanswered. Recent work by \citet{sal09a} is an example of how the kinematics predicted by the four different scenarios---in particular, the distribution of stars' orbital eccentricities---can be used to test the various scenarios. This approach has been taken observationally by \citet{die10} and \citet{wil11} using stars from the SDSS and the RAdial Velocity Experiment (RAVE; \citealt{ste06}), respectively. Both studies disfavor the minor merger scenarios, as they do not observe enough stars with high orbital eccentricities. Numerical simulations by \citet{di-11}, however, suggest that the observed eccentricity distribution can be obtained in the minor merger scenario given different orbital parameters and satellite properties than those used in the \citet{sal09a} analysis. The conflicting interpretations show the need for other observational constraints. In this paper we examine the radial metallicity gradient $\Delta{\rm[Fe/H]}/\Delta R$, as a function of height above the Galactic plane $|Z|$, in order to further distinguish between different thick disk formation scenarios.

For example, if thick disk stars originated in a thin disk and were subsequently heated by a minor merger as in scenario 1, then the observed radial metallicity gradient depends on the amount of mixing in the radial direction. If most of the heating occurs in the vertical direction, the thick disk will have the same metallicity gradient as the initial thin disk. Simulations, however, show that there can be substantial heating in the radial direction (e.g., \citealt{hay06,kaz09,bir11}), which suggests that the gradient may be more shallow or flat than expected. If thick disk stars originated outside of the Galaxy and were deposited in the thick disk via accretion events as in scenario 2, then the metallicity distribution may exhibit clumpiness. The simulations of \citet{aba03} showed that a single disrupted satellite roughly ends up in a torus of stars; several disrupted satellites would make up the thick disk by contributing stars of different metallicities at different radii. 

If thick disk stars originated in a turbulent gas disk at high redshift as in scenario 3, the short timescale for star formation makes the thick disk chemically homogeneous, with no metallicity gradient \citep{bro05,bou09}. If thick disk stars originated in a thin disk and were pushed to larger radii through radial migration as in scenario 4, then the original gradient would become washed out into a shallow or nonexistent radial metallicity gradient \citep{ros08b,san09}. The presence of a gradient, then, would rule out chemical homogeneity (scenario 3), while the strength of the gradient would constrain the amount of disk heating by minor mergers (scenario 1) and radial migration (scenario 4). Examining old disk stars in a large volume, beyond the solar neighborhood, will allow one to distinguish between the various scenarios.

In contrast to many previous studies, we do not assign our stars to a thin disk or thick disk component. For samples of nearby stars in the solar neighborhood, this division is often done by assuming that the thick disk has a larger velocity dispersion and a slower mean rotation (i.e., thick disk stars are kinematically hot), as in the cases of \citet{ben03} and \citet{ven04}. Another method of separating thin and thick disk stars is by their chemistry, as \citet{lee11b} do using [Fe/H] and [$\alpha$/Fe]; they favor this type of division because a star's composition is less likely to change than its spatial location or kinematics. \citet{sch09b} have used mock observations to show that separating thin and thick disk stars using chemistry versus kinematics yields samples with different properties.\footnote{\footnotesize For example, a star assigned to the thin disk using chemical criteria may be assigned to the thick disk using kinematic criteria if it belongs to the inner disk and is in the tail of the rotational velocity distribution. According to the \citet{sch09b} model, this explains the tail of thick disk stars with high [Fe/H] and why most of the thick disk stars in the \citet{ben03} and \citet{ven04} samples are located at radii within the solar circle.} In addition, whether the thin and thick disks are truly distinct components is still an open question, with some studies arguing that the two components arise from a smooth correlation between chemical and kinematic properties (see discussions by \citealt{hay08} and \citealt{ive08}). 

Because our sample is not restricted to the solar neighborhood, we can compare the stellar populations of the thin and thick disks based on stars' locations instead of their kinematics or chemistry, which allows us to avoid assuming a specific model for the Milky Way disk. In this paper, we do not assign individual stars to the thin or thick disk. Instead, we will use the term \textit{thick disk} to refer to stars that are currently found at large distances from the plane on their orbits; the term \textit{thin disk} refers to stars that are found close to the Galactic mid-plane. Based on double-exponential fits to the vertical scale heights of the stellar density distribution in the disk (e.g., \citealp{jur08,dej10}) we expect the thick disk to be the dominant population above $|Z|\sim1.0$ kpc.

Previous analyses of samples outside of the solar neighborhood have found no radial metallicity gradient at vertical heights $|Z| > 1.0$ kpc \citep{all06,jur08}. Our sample is complementary because our lines of sight are located at relatively low Galactic latitude and we can directly compare the radial metallicity gradients of the thin and thick disks (up to $|Z|=1.5$ kpc) using the same sample. In addition, we can explore whether the reported discontinuity in the outer disk is a purely \textit{radial} trend or if a \textit{vertical} trend is contributing to the observed flattening of the gradient at large $R$.

The paper is organized as follows: The sample selection and data are described in \S\ref{data}. We then describe our methods of determining distances and correcting for the selection function in \S\S\ref{distances} and \ref{weights}, respectively. Our gradient measurements are presented in \S\ref{results}. Error analysis is presented in \S\ref{errors}. We discuss the results in \S\ref{discussion} and conclude with a summary in \S\ref{summary}. For readers who are only interested in the results, we recommend skipping \S\S\ref{weights} and \ref{errors}. Further description of our weighting scheme, introduced in \S\ref{weights}, is provided in the Appendix. Throughout our analysis we adopt the Galactocentric radius of the Sun, $R_{\rm GC, \odot}=8.0$ kpc.

\section{Data}\label{data}
\subsection{Sample Selection}\label{targets}
We measure the metallicity gradient of old main sequence turnoff (MSTO) stars in low Galactic latitude fields from the SEGUE survey (\citealt{yan09,aih11,eis11}; Rockosi et al. 2011, in preparation). These stars allow us to reach the largest distances probed by main sequence stars within the fixed magnitude limits of the survey. The data were obtained using the same telescope \citep{gun06}, camera \citep{gun98}, and filter system \citep{fuk96} as the SDSS. The old MSTO is selected to be in the blue part of the color-magnitude diagram (CMD), as described in detail below, and can be identified using the SDSS Data Release 7 (DR7; \citealt{aba09}) version of the Catalog Archive Server\footnote{\footnotesize http://casjobs.sdss.org/CasJobs/} as targets with \texttt{sspParams.zbclass = STAR, SpecObjAll.primTarget = 2048}, and \texttt{PlateX.programName = seglow\%}. We also require that there are no repeat observations so that each star is only counted once. An equivalent query in Data Release 8\footnote{\footnotesize http://skyservice.pha.jhu.edu/casjobs/} (DR8; \citealt{aih11}) is \texttt{SpecObjAll.class = STAR, SpecObjAll.primTarget = 2048, PlateX.programName = seglow\%}, and \texttt{PlateX.isPrimary = 1}, where the last requirement removes repeat observations.

The \texttt{programName} qualifier selects our targets from a subset of 22 SEGUE plug-plates that comprise the \textquotedblleft Low-Latitude" pointings, which are restricted to Galactic latitudes $8^{\circ} < |b| < 16^{\circ}$ (see \S3.15 of \citealt{yan09}). These lines of sight are high enough to avoid the young star-forming disk, as well as the regions with the most crowding and highest reddening, but also sufficiently low that they have a long sightline through the disk. The lines of sight fall into roughly two groups in Galactic longitude: seven at $50^{\circ} < l < 110^{\circ}$ and another four toward the anticenter, $170^{\circ} < l < 210^{\circ}$. 

Each plate covers 7 square degrees on the sky with targets in the magnitude range $16 < g < 20$, where the magnitudes have not been corrected for extinction. In this paper, we will refer to any reddening- or extinction-corrected magnitudes and colors with subscripts $g_{\rm SFD}$ and $g_0$, for corrections derived from \citet[hereafter \citetalias{sch98}]{sch98} and isochrone fitting (see \S\ref{distances}), respectively. Table~\ref{los} lists the properties of the 11 lines of sight (two plates per pointing) included in our sample, ordered by the median extinction $E(B-V)$, obtained from \citetalias{sch98}. For the total sample, $E(B-V)$ varies between 0.05 and 1.07 magnitudes. On average, there were 600-700 spectra obtained per line of sight.

\begin{table*}
\caption{Properties for 11 Lines of Sight.}
\centering
\scriptsize\begin{tabular}{llrrrrcc}
\hline
\multicolumn{2}{c}{Plates} & \multicolumn{1}{c}{$RA$ ($^{\circ}$)} & \multicolumn{1}{c}{$DEC$ ($^{\circ}$)} & \multicolumn{1}{c}{$l$ ($^{\circ}$)} & \multicolumn{1}{c}{$b$ ($^{\circ}$)} & $E(B-V)^{\rm a}$ & N$_{\rm spectra}$\\
\hline
2712 & 2727 & 105.6 & 12.4 & 203.0 & 8.0 & 0.09 & 744\\
2536 & 2544 & 286.7 & 39.1 & 70.0 & 14.0 & 0.15 & 661\\
2534 & 2542 & 277.6 & 21.3 & 50.0 & 14.0 & 0.17 & 672\\
2554 & 2564 & 303.0 & 60.0 & 94.0 & 14.0 & 0.19 & 734\\
2678 & 2696 & 98.1 & 26.7 & 187.0 & 8.0 & 0.24 & 766\\
2556 & 2566 & 330.2 & 45.1 & 94.0 & -8.0 & 0.31 & 728\\
2668 & 2672 & 79.5 & 16.6 & 187.0 & -12.0 & 0.33 & 830\\
2681 & 2699 & 71.5 & 22.0 & 178.0 & -15.0 & 0.41 & 758\\
2537 & 2545 & 334.2 & 69.4 & 110.0 & 10.5 & 0.49 & 708\\
2538 & 2546 & 323.1 & 73.6 & 110.0 & 16.0 & 0.65 & 716\\
2555 & 2565 & 312.4 & 56.6 & 94.0 & 8.0 & 0.82 & 511\\
\hline
\label{los}
\end{tabular}
\\$^{\rm a}$ Median value for spectra in line of sight using values from \citetalias{sch98}.
\end{table*}

For most of the SEGUE survey, which was at high Galactic latitude, targets are identified as MSTO stars based on their $(u-g)_{\rm SFD}$ and $(g-r)_{\rm SFD}$ colors (see \citealt{yan09}). In $ugr$ color space, it is possible to separate the MSTO stars from metal-poor halo stars because of the large ultraviolet excess of metal-poor stars. In the low latitude pointings, however, two issues arise. First, the $u$-band magnitudes and their uncertainties are unreliable due to the large extinction in these regions. Second, it is impossible to use a single constant selection in $(g-r)_{\rm SFD}$ that will yield the same stellar population in every line of sight because the reddening in these fields is, on average, much higher and more variable than in the high latitude fields. For these low latitude fields, we use a targeting procedure that is more robust to reddening, which will reliably choose the stars at the blue edge of the CMD. Starting with the photometric objects identified as stars in the imaging, this selection procedure is as follows: 

\begin{enumerate}
\item{We remove all stars with $g > 20$ and $i < 14.2$ (using magnitudes \textit{uncorrected for extinction}) to ensure that the targets are bright enough for high-quality spectroscopy in the expected exposure time.}
\item{The 7 square degree area of each plate is large enough that the extinction can be highly variable across the plate, and there are always many more targets than fibers available. We remove the regions of highest extinction from consideration to maximize the number of useful spectra. For each half of the plate, we calculate the 75th percentile of the $E(B-V)$ distribution using the total line of sight extinction from \citetalias{sch98}. This procedure is done for each half independently to ensure that the targets are approximately evenly distributed over the plates, since the reach of the fibers is only about half of the plate diameter.}
\item{We remove all objects with $E(B-V)$ larger than the higher of the two 75th percentile values. Taking the higher value ensures that there are enough usable targets on each half of the plate to fill all the fibers given their limited reach across the plate. This should not bias the sample, as we do not expect that the objects behind more extinction are intrinsically any different from those that are unobscured. This is especially true for distant objects that are located far behind the dust. \textit{The magnitudes used throughout the rest of this procedure are corrected using the \citetalias{sch98} extinction values.} The \citetalias{sch98} extinction was applied so that stars in the same approximate luminosity range were targeted in each line of sight despite the large variation in extinction among the different lines of sight.}
\item{We examine the $(g-r)_{\rm SFD}$ distribution in bins of $g_{\rm SFD}$, each one magnitude wide. For each distribution, we find the peak, which is the $(g-r)_{\rm SFD}$ color of the MSTO in each $g$-magnitude bin. In addition, we determine the half maximum on the blue side; this is $(g-r)_{\rm half-max}$.}
\item{The red cut for each bin is defined as $(g-r)_{\rm cut}\equiv(g-r)_{\rm half-max}+0.25$. We fit a line to $(g-r)_{\rm cut}$ as a function of $g_{\rm bin}$, where $g_{\rm bin}$ is the mean $g_{\rm SFD}$ of all the stars in each bin.}
\item{All stars on the blue side of the line are defined as candidate spectroscopic targets with equal probability of being selected. Targets are randomly chosen from the resulting candidate list.}
\end{enumerate}

Though the $(g-r)_{\rm SFD}$ color of the population may change from field to field because of varying amounts of extinction, the identification of the MSTO stars as the bluer population holds in all lines of sight. As a result, this method is more robust to reddening than the standard color cuts for normal SEGUE plates. Halo contamination is expected to be low in these plates, compared with the higher latitude pointings (see further discussion in \S\ref{halo}). The color cut, however, will bias our sample against metal-rich stars, which have redder colors. The severity of the bias depends on how much the MSTO color changes with metallicity, which in turn depends on the ages of stars at each metallicity (see further discussion in \S\S\ref{metalbias} and \ref{gradient_errors}).

\begin{figure*}[!ht]
\epsscale{0.55}
\plotone{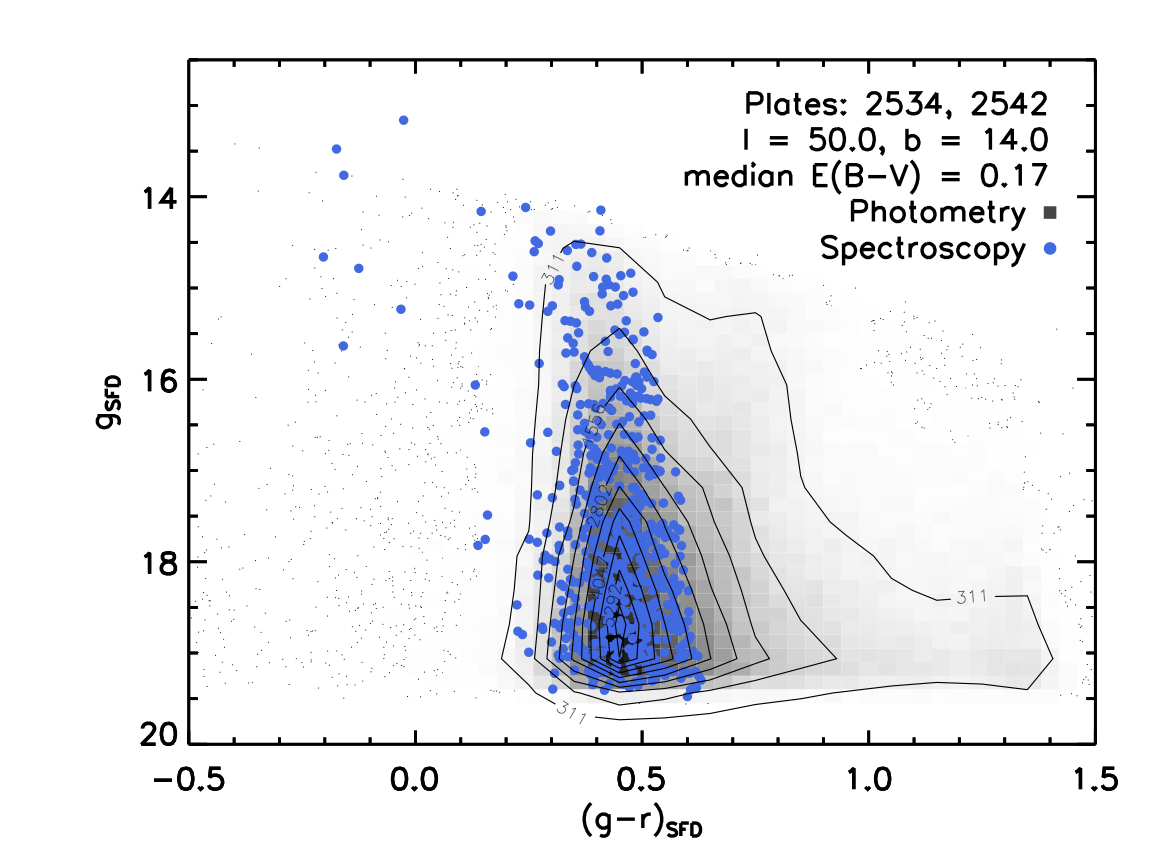}
\plotone{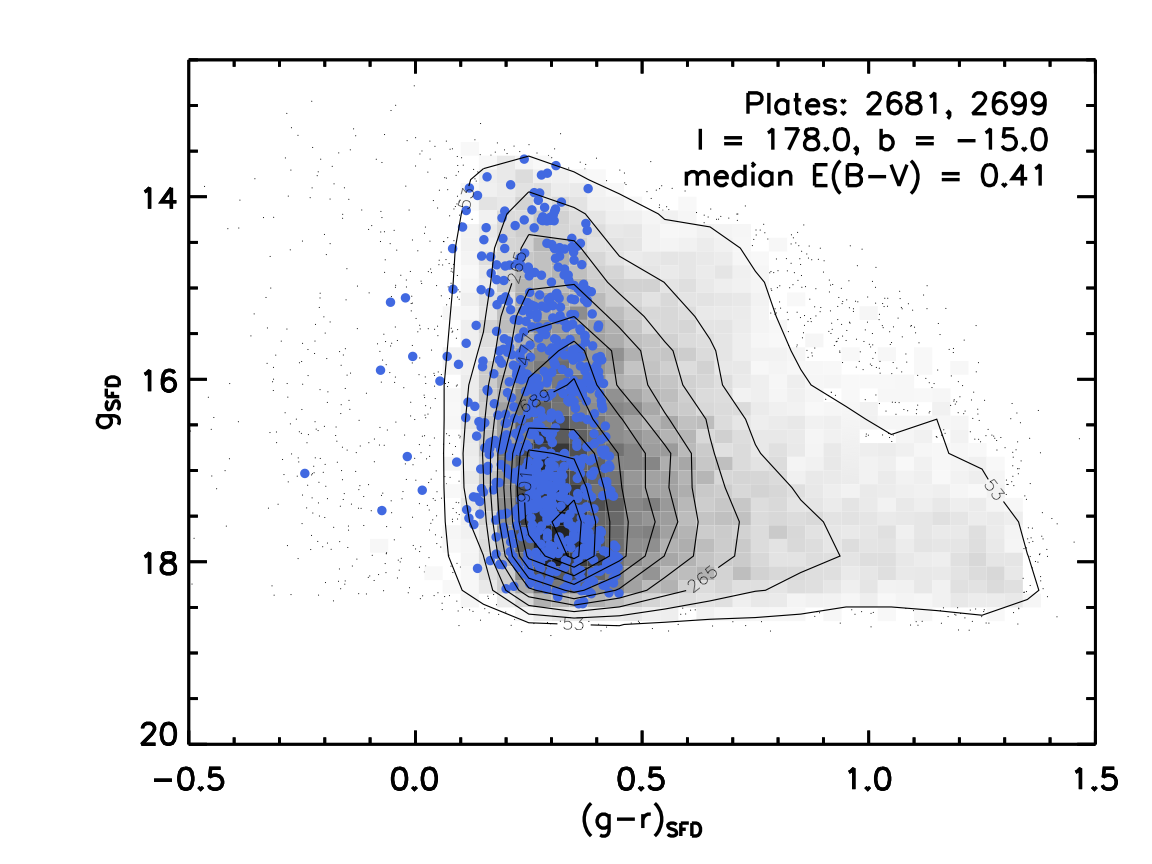}
\caption{Main sequence turnoff (MSTO) selection. Spectroscopic targets (blue circles) are a randomly-selected subset of the MSTO stars identified using the photometry of all objects in the field (grayscale and contours). The contour labels indicate the number of stars per 0.375 by 0.1 magnitude $g_{\rm SFD}$-$(g-r)_{\rm SFD}$ bin. The results are shown for two lines of sight with low and high extinction (left and right, respectively), as measured by \citetalias{sch98}.}
\label{cmd_selection}
\end{figure*}

Figure~\ref{cmd_selection} shows the results of the procedure outlined above for two lines of sight with low and high extinction (median $E(B-V)=$ 0.17 and 0.41, respectively). The density of objects identified as stars in the photometry are plotted in grayscale and contours, while MSTO stars in our spectroscopic sample are plotted as blue circles. All photometric objects bluer than the red limit of the spectroscopic sample were considered as candidates for spectroscopy, but only a randomly selected subset of those were actually observed. In the 22 low latitude plates, spectra of 7828 MSTO stars were taken. We keep the targets with good photometry and spectra with $S/N>10$ per pixel, where each pixel corresponds to $\sim1\AA$. We remove any stars which have large discrepancies ($>0.1$ mag) in $g-r$ color between different photometric reductions. We also remove rare blue stars with $(g-r)_{\rm SFD} < -0.25$. The resulting sample contains 7655 spectra with a mean $S/N\sim30$ per pixel.

\subsection{SEGUE Stellar Parameter Pipeline: Accuracy in Regions of High Extinction}\label{sspp}
The SEGUE Stellar Parameter Pipeline (SSPP; \citealp{lee08a}) estimates the effective temperature $T_{\rm eff}$, surface gravity log $g$, and metallicity [Fe/H] for each spectroscopic target in the survey. We use stellar parameters from the version of the SSPP used for Data Release 8, which includes improved [Fe/H] estimates at both high and low metallicities \citep{smo11}. The SSPP has been extensively tested using globular and open clusters, where true cluster members are identified using their metallicities and radial velocities \citep{lee08b,smo11}. 

\begin{table}
\caption{Cluster Metallicities Measured by the SSPP.}
\centering
\scriptsize\begin{tabular}{lrcc}
\hline
Cluster$^{\rm a}$ & \multicolumn{1}{c}{[Fe/H]$_{\rm Lit}$} & \multicolumn{1}{c}{[Fe/H]$_{\rm SSPP}$} & \multicolumn{1}{c}{[Fe/H]$_{\rm Lit}-$[Fe/H]$_{\rm SSPP}$}\\
\hline
M3 &  -1.50\phantom{$^{\rm b}$} &  -1.43 &  -0.07\\
M71 &  -0.82\phantom{$^{\rm b}$} &  -0.74 &  -0.08\\
NGC2158 &  -0.25\phantom{$^{\rm b}$} &  -0.29 &   +0.04\\
NGC2420 &  -0.20$^{\rm b}$ &  -0.31 &   +0.11\\
M35 &  -0.16\phantom{$^{\rm b}$} &  -0.24 &   +0.08\\
M67 &   +0.05$^{\rm c}$ &   +0.00 &   +0.05\\
NGC6791 &   +0.30\phantom{$^{\rm b}$} &   +0.29 &   +0.01\\
\hline
\label{cluster_dr8}
\end{tabular}
\\$^{\rm a}$ Cluster data are also presented in \citet{smo11}, with the exceptions of NGC2420 and M67. A full comparison of the SSPP for all data in the SEGUE cluster samples will be presented in Rockosi et al. (2012, in prep).
\\$^{\rm b}$ \citet{jac11}.
\\$^{\rm c}$ \citet{ran06}.
\end{table}

We verify that these results hold for cluster members in our temperature and surface gravity range ($5000 < T_{\rm eff} < 7000$ K, log $g > 3.3$). Table~\ref{cluster_dr8} shows the comparison between the literature values (column 2) and the SSPP for cluster members in our temperature and surface gravity range (columns 3-4). The offsets in [Fe/H] between the literature values and the SSPP are small (within 0.1 dex), and we see no trends with $T_{\rm eff}$, [Fe/H], and $S/N$. These tests show that we can reliably measure trends in [Fe/H] throughout our entire sample volume, and that the absolute values of the metallicities presented in the paper are accurate to 0.1 dex or better.

Each parameter is estimated using multiple methods: 11 for $T_{\rm eff}$, 10 for log$g$, and 12 for [Fe/H]. For $T_{\rm eff}$ in particular, these methods include spectral fitting and $\chi^2$-minimization using grids of synthetic spectra (ki13, k24, NGS1), measuring line indices (WBG, HA24, HD24), neural networks using training sets of both real and synthetic spectra (ANNRR, ANNSR), and $g-r$ color predictions (${\rm T_K, T_G, T_I}$). See \citet{lee08a} for complete details on each of these individual methods. Some methods use only the spectra as input, some use only the photometry, and some use both. The individual estimates of each parameter are averaged to obtain a final adopted value. Each individual estimate is valid for some range of $(g-r)_{\rm SFD}$ color and $S/N$, which determines whether it will be included in the final average. A more complicated decision tree is used for [Fe/H] and is described in detail in the Appendix of \citet{smo11}. We have 7605 stars that have good stellar parameters; their temperatures fall in the range $5000<T_{\rm eff}<7000$ K, making them F and G dwarfs.

Before using the SSPP parameters, we test whether the parameter estimates are affected by high Galactic extinction. The SSPP was designed to analyze the normal SDSS and SEGUE data at high Galactic latitude and uses photometry that has been corrected for extinction using the reddening maps of \citetalias{sch98}. These extinction values reflect the total line-of-sight extinction, which means that the colors of less distant stars will be overcorrected; they will be too blue. This effect is likely to result in a systematic error in the parameters estimated by those methods in the SSPP that use the photometry.

\begin{figure*}[!ht]
\epsscale{1.1}
\plotone{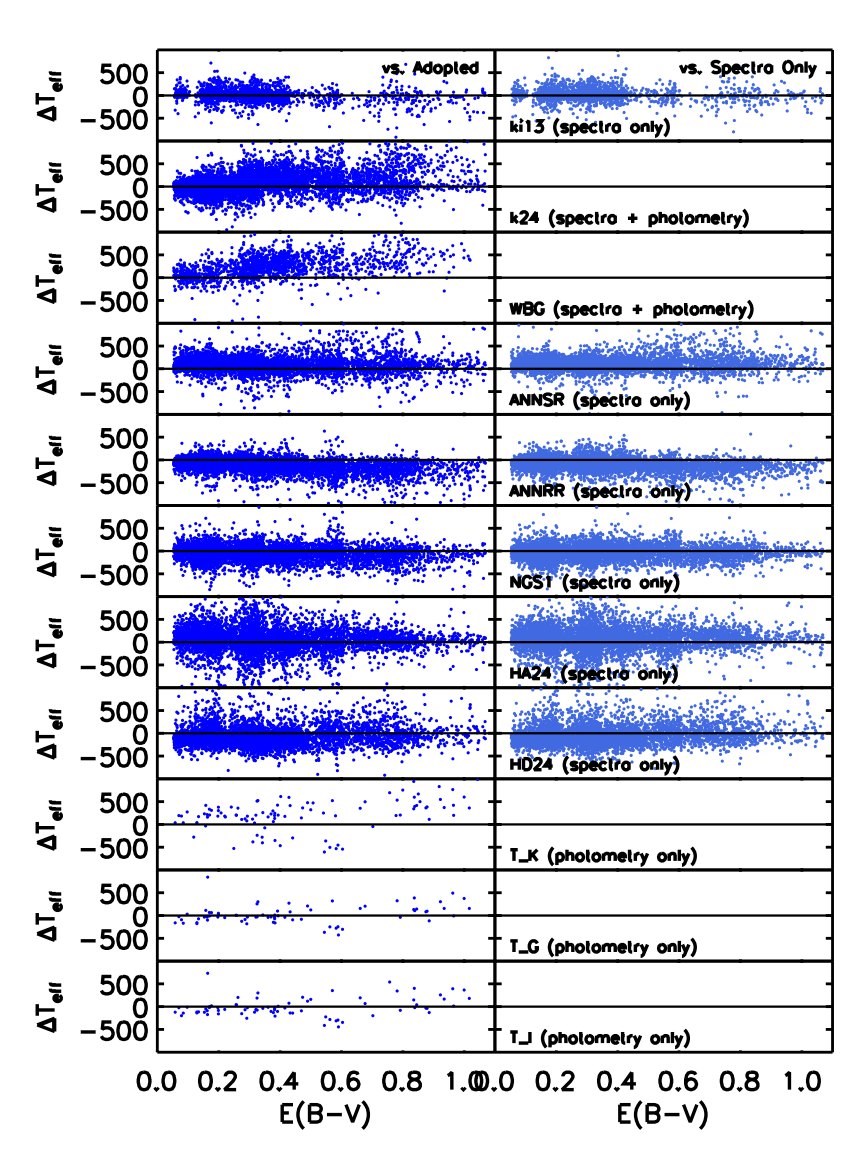}
\caption{Left panels: Differences between individual SSPP temperature estimates and the adopted values as a function of \citetalias{sch98} extinction $E(B-V)$. For estimates that include photometry (k24, WBG, ${\rm T_K, T_G, T_I}$), the deviation from the adopted value increases with extinction, indicating that the temperatures are overestimated due to overcorrection when using the \citetalias{sch98} extinction values. Right panels: Differences between Individual SSPP temperature estimates and the spectra-only values as a function of extinction. The trend is no longer evident in the spectra-only estimates, indicating that the spectra-only temperature is more reliable for highly extincted objects.}
\label{ebv_teff}
\end{figure*}

If this overcorrection affects our sample, we expect estimates that use the photometry to be systematically different when the extinction is high. The left panels of Figure~\ref{ebv_teff} show the discrepancy between the individual $T_{\rm eff}$ estimates and the adopted value as a function of $E(B-V)$. Whether the estimate includes the photometry is indicated in the bottom left of each righthand panel. We have only plotted estimates that were accepted by the SSPP (i.e., the target falls in the $(g-r)_{\rm SFD}$ or $S/N$ range in which the particular method is reliable). As expected, the photometry-dependent estimates of $T_{\rm eff}$ (k24, WBG, ${\rm T_K, T_G, T_I}$) are systematically higher (i.e., the color is bluer), compared to the adopted value, for the highest values of extinction.

To remove this effect, we calculate new averaged values of $T_{\rm eff}$ using only the estimates from methods that exclude the photometry (ki13, ANNSR, ANNRR, NGS1, HA24, HD24). The right panels of Figure~\ref{ebv_teff} show the discrepancy between the individual $T_{\rm eff}$ estimates and the newly calculated spectra-only value as a function of $E(B-V)$. In contrast to the left panels, there is no clear trend in $\Delta T_{\rm eff}$ with extinction for the individual spectra-only estimates, which suggests that our new spectra-only average of the temperature is more reliable. At $E(B-V) > 0.8$, the median discrepancy between the new spectra-only estimate and the original adopted value is about $100-200$K, which is comparable to the expected errors of the SSPP $T_{\rm eff}$. We show in \S\ref{distanceerrors} that this could amount to a systematic error in the distance of $\sim 20-25\%$. For the remainder of the paper, $T_{\rm eff}$ will refer to the spectra-only value. 

No trend with extinction is observed in the photometry-dependent estimates of [Fe/H] (k24, WBG, CaIIK2, CaIIK3, ACF, CaII), so we keep the adopted values. This provides a more robust result, as the adopted value is an average of a larger number of estimators.

\section{Distances}\label{distances}
To calculate distances to each target, we use the spectra-only $T_{\rm eff}$ and SSPP [Fe/H], plus the theoretical isochrones of \citet{an09}\footnote{\footnotesize{http://www.astronomy.ohio-state.edu/iso/sdss.html}}, which have been shown to be good matches to $ugriz$ cluster fiducials. We do not make use of the SSPP log $g$ estimate, as it is relatively inaccurate near the turnoff, where the expected range of gravities is small compared to the errors. Twelve sets of 
\citet{an09} isochrones at metallicities in the range $-3.0 < {\rm [Fe/H]} < +0.4$ are available, with each set having a range of ages up to 15.8 Gyr. The distance uncertainties are discussed in \S\ref{distanceerrors}.

We assign all the stars in the sample to the isochrone with the closest metallicity. We then find the mean temperature for the stars in each metallicity bin by fitting a Gaussian to the distribution of effective temperatures. We identify the age of the isochrone with the turnoff temperature closest to the measured mean temperature; we refer to this as the \textit{turnoff age of the mean temperature}, or TAMT, for each metallicity. Targets hotter than the mean temperature cannot be placed on the TAMT isochrone, so younger isochrones must be used; these targets are assigned to the oldest possible age (i.e., the oldest isochrone where the target is cooler than the turnoff). 

\begin{figure}[!ht]
\epsscale{1.1}
\plotone{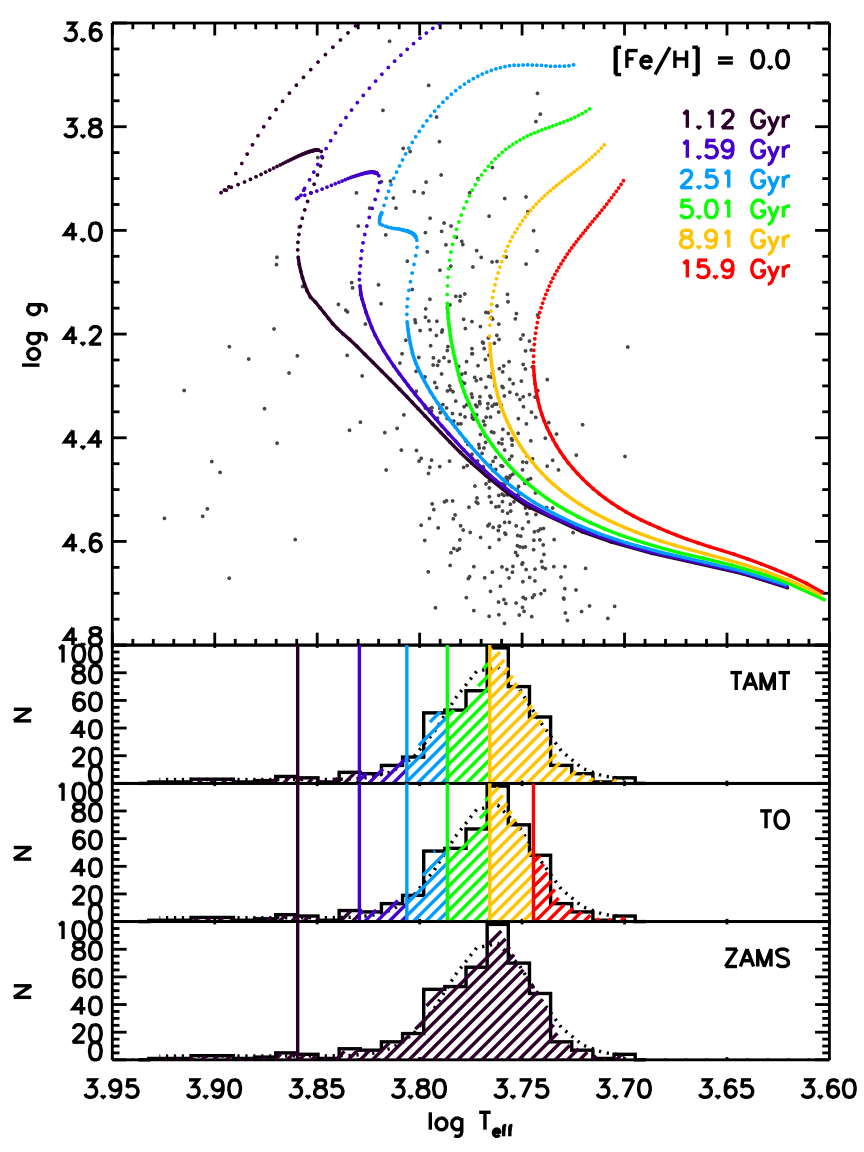}
\caption{Determination of isochrone ages for solar metallicity targets ($-0.05 < $ [Fe/H] $ < 0.05$, gray dots). Isochrones are shown in the top panel, with the temperature distribution in the bottom three panels. In practice, the set of isochrones used at each metallicity includes all available ages, but for clarity we show only six --- the oldest, youngest, and those located at the mean and 1-, 2-, and 3-$\sigma$ values of the temperature distribution. The colors in the lower panels indicate which isochrones in the top panel are used to calculate distances for those targets. We show how ages are assigned to our targets making three possible age assumptions: Turnoff Age of the Mean Temperature (TAMT), Turnoff (TO), and Zero-Age Main Sequence (ZAMS). TAMT: For each metallicity, most targets are assigned to a single age (the TAMT) at which the turnoff temperature is closest to the measured mean of the temperature distribution of the SEGUE targets. Targets hotter than the mean are assigned the oldest possible age (i.e., the oldest isochrone where the target is cooler than the turnoff). TO: All targets are assigned to the oldest possible age, even those cooler than the mean of the temperature distribution. ZAMS: All targets are assigned to the youngest possible age. The TAMT assumption is used in our analysis, while the TO and ZAMS assumptions are used to test the accuracy of our distance estimates (see \S\ref{distanceerrors}).}
\label{isochrone_ages}
\end{figure}

Figure~\ref{isochrone_ages} shows a schematic picture of how ages are assigned to targets depending on their effective temperatures. The top panel shows six representative isochrones at solar metallicity, while the bottom three panels show how stars are assigned to these isochrones based on their place in the $T_{\rm eff}$ distribution. The procedure described above is shown in the panel labeled \textquotedblleft TAMT" while the panels labeled \textquotedblleft TO" and \textquotedblleft ZAMS" show two other age assumptions that we use to estimate our distance errors (\S\ref{distanceerrors}). Table~\ref{meanages} lists the TAMT determined by finding the mean temperature for each metallicity bin for the 7605 stars in our sample. 

\begin{table*}
\caption{Mean Isochrone Ages as a Function of [Fe/H].}
\centering
\scriptsize\begin{tabular}{ccccrrr}
\hline
[Fe/H] & [$\alpha$/Fe] & log $\langle T_{\rm eff}\rangle$ & $\langle T_{\rm eff} \rangle$ (K) & \multicolumn{1}{c}{log $\langle age \rangle$} & \multicolumn{1}{c}{$\langle age \rangle$ (Gyr)} & \multicolumn{1}{c}{N$_{\rm obj}$}\\
\hline
\phantom{-3.00 $<$} [Fe/H] $<$ -2.50 & +0.4 & 3.768 & 5864 & 10.2 & 15.8 & 9\\
-2.50 $<$ [Fe/H] $<$ -0.75& +0.3 & 3.768 & 5864 & 10.20 & 15.8 & 880\\
-0.75 $<$ [Fe/H] $<$ -0.40 & +0.2 & 3.775 & 5958 & 10.05 & 11.2 & 2695\\
-0.40 $<$ [Fe/H] $<$ -0.25 & \phantom{+}0.0 & 3.772 & 5912 & 10.05 & 11.2 & 1527\\
-0.25 $<$ [Fe/H] $<$ -0.15 & \phantom{+}0.0 & 3.772 & 5912 & 10.00 & 10.0 & 857\\
-0.15 $<$ [Fe/H] $<$ -0.05 & \phantom{+}0.0 & 3.772 & 5912 & 9.95 & 8.9 & 732\\
-0.05 $<$ [Fe/H] $<$ +0.05 & \phantom{+}0.0 & 3.763 & 5796 & 10.00 & 10.0 & 480\\
+0.05 $<$ [Fe/H] $<$ +0.15 & \phantom{+}0.0 & 3.763 & 5796 & 9.95 & 8.9 & 229\\
+0.15 $<$ [Fe/H] $<$ +0.30 & \phantom{+}0.0 & 3.763 & 5796 & 9.85 & 7.1 & 142\\
+0.30 $<$ [Fe/H] \phantom{$<$ +0.50} &  \phantom{+}0.0 & 3.763 & 5796 & 9.70 & 5.0 & 54\\
\hline
\label{meanages}
\end{tabular}
\end{table*}

Once we have assigned a given target to an isochrone with a particular age and metallicity, we use the isochrone to obtain the predicted $g-r$ color for the target's spectra-only $T_{\rm eff}$ by linearly interpolating on the isochrone in temperature-color space. A comparison of the predicted and observed $g-r$ gives an estimate of the \textit{isochrone extinction} in $g-r$, which is also used to determine the extinction in the $g$-band. The isochrone extinction is an improvement over the \citetalias{sch98} values because it does not assume that the target lies behind all of the dust in the line of sight. We step through this procedure for the two isochrones with the nearest values of [Fe/H] and then linearly interpolate to find the predicted apparent and absolute magnitudes. The apparent magnitude, now corrected using the isochrone extinction, along with the predicted absolute magnitude in the $g$-band yield the distance.

\begin{figure*}[!ht]
\epsscale{0.55}
\plotone{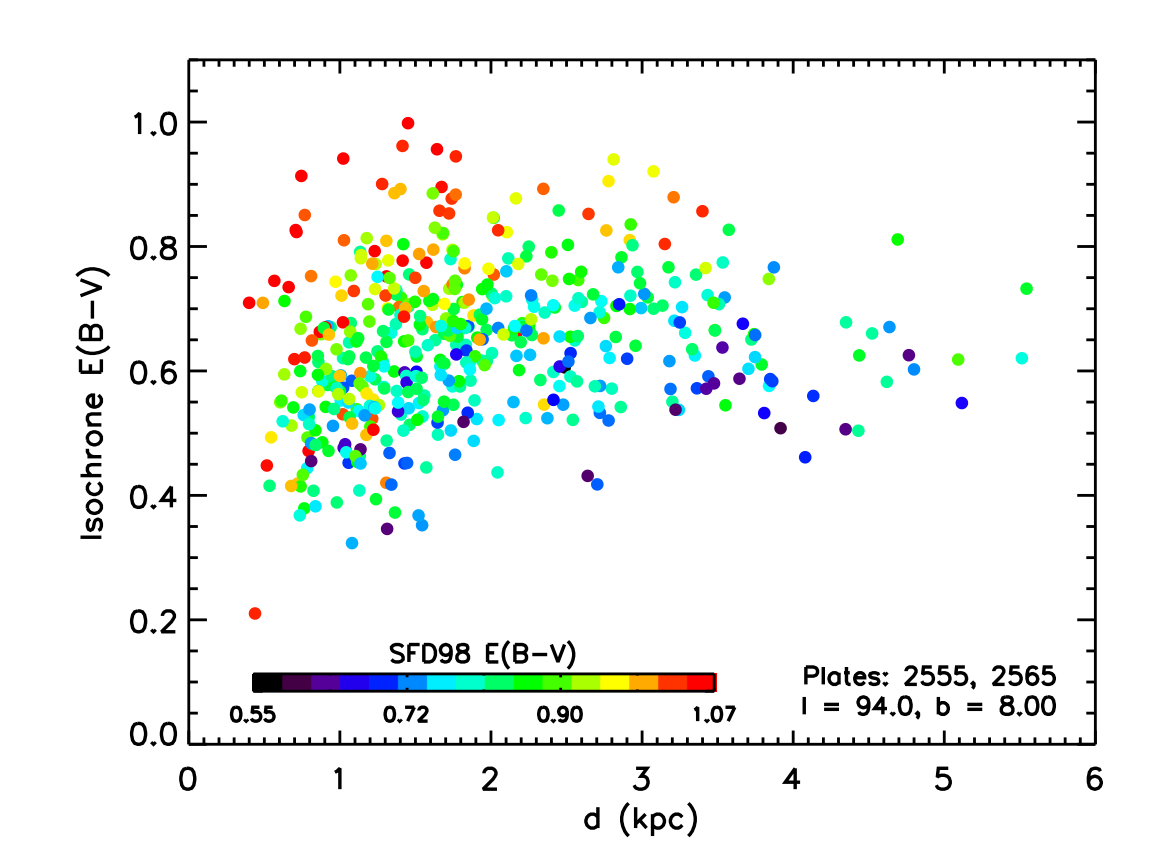}
\plotone{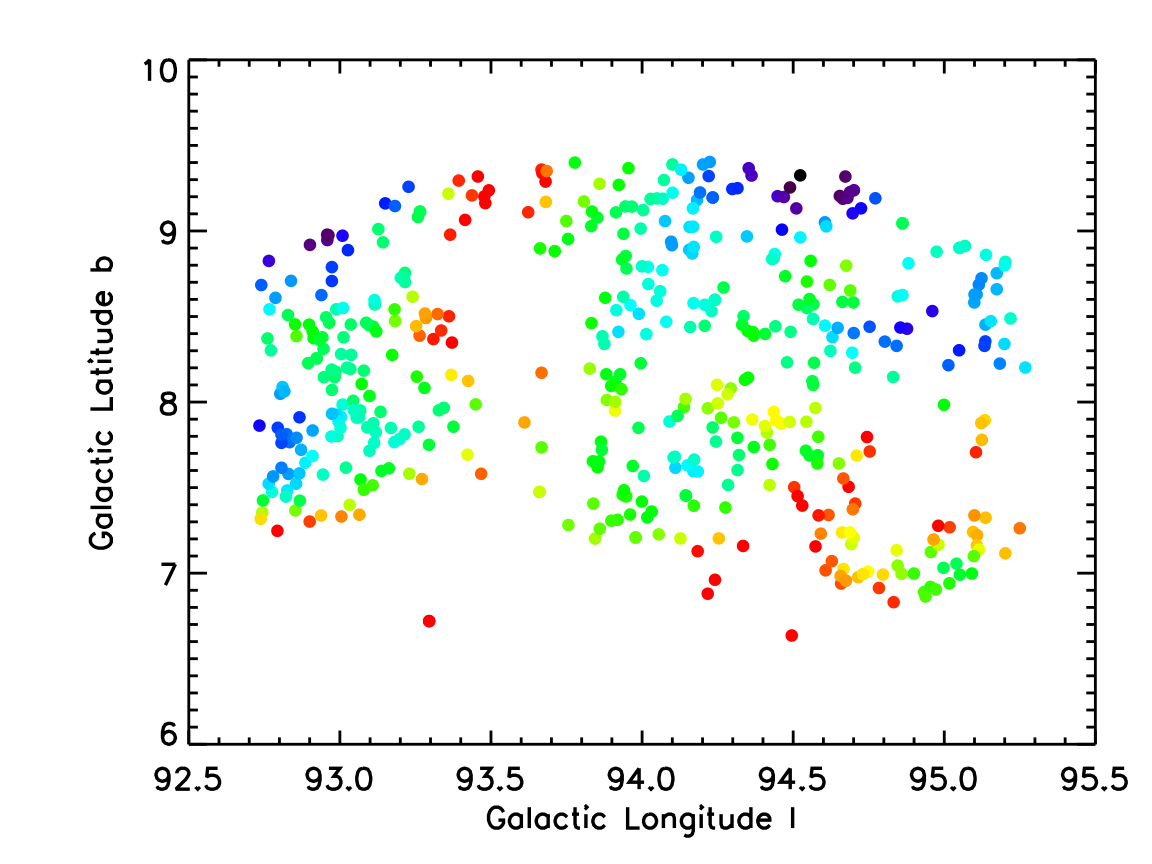}
\caption{Estimates of the extinction derived from isochrone fitting give a picture of the dust distribution along a given line of sight. The left panel shows the isochrone extinction as a function of derived distance $d$ from the Sun. At small $d$ the isochrone extinction is smaller than the \citetalias{sch98} value (indicated by symbol color), but at large $d$ it asymptotes to the \citetalias{sch98} value. For this line of sight, which has the highest median $E(B-V)$ of our 11 fields, most of the dust is found within 2 kpc of the Sun. The scatter in isochrone extinction for a given value of \citetalias{sch98} extinction may reflect the patchiness of the dust on the plane of the sky. This effect is seen in the right panel, which shows multiple separate patches that have the same \citetalias{sch98} extinction values, but may have different distributions of dust along the line of sight. The color coding in the right panel is the same as in the left.}
\label{d_gmrext}
\end{figure*}

In addition to being used in the distance calculation, the isochrone extinction provides information about the dust distribution along different lines of sight in the field. The left panel of Figure~\ref{d_gmrext} shows the isochrone extinction as a function of the derived distance for one line of sight, with the color indicating the \citetalias{sch98} value. There is good general agreement between the two extinction estimates, especially on a relative scale. Quantitatively, the agreement is poorest for small distances where the targets are in front of some of the dust, but asymptotes to better agreement at large distances. This is consistent with the idea that the \citetalias{sch98} values are overestimates because they include all of the dust in the given line of sight. For the line of sight shown, most of the dust is located within 2 kpc of the Sun, beyond which the isochrone extinction is approximately constant as a function of distance.

The scatter in the left panel of Figure~\ref{d_gmrext} may be explained by the patchiness of the extinction on the plane of the sky. The right panel of Figure~\ref{d_gmrext} shows that the distribution of $E(B-V)$ varies on small scales. The scatter in the left panel is expected if each region of high extinction has a different dust distribution along the line of sight. The agreement between the isochrone extinction and the \citetalias{sch98} values provides a sanity check which indicates that our isochrone extinction estimates are reliable.

Sixty-two stars in our sample (0.8$\%$) end up with negative values of isochrone extinction. Three kinds of stars fall into this category: (1) 17 stars are outliers with very blue $g-r$ colors. These stars do not contribute to the final measurement; after we apply our weighting scheme they receive a CMD weight of zero (see \S\ref{weights}). (2) 7 stars show large changes in $g-r$ color between different photometric reductions and are likely to be catastrophic errors, possibly due to blending in the relatively crowded, low-latitude fields. (3) The remaining 38 stars tend to be faint, and likely have negative reddening because the errors in their temperatures cause their predicted $g-r$ colors to be redder than their observed colors. We only see these stars in fields where the reddening is low; in fields where reddening is higher, the observed $g-r$ is much redder than the predicted $g-r$, so the temperature errors are not large enough to cause stars to have negative reddening. We expect just as many objects to have temperature errors that cause their predicted $g-r$ colors to be bluer, and we leave all of these objects in our sample, which should give us a more symmetric distribution of errors in our distances.

\begin{figure}[!ht]
\epsscale{1.1}
\plotone{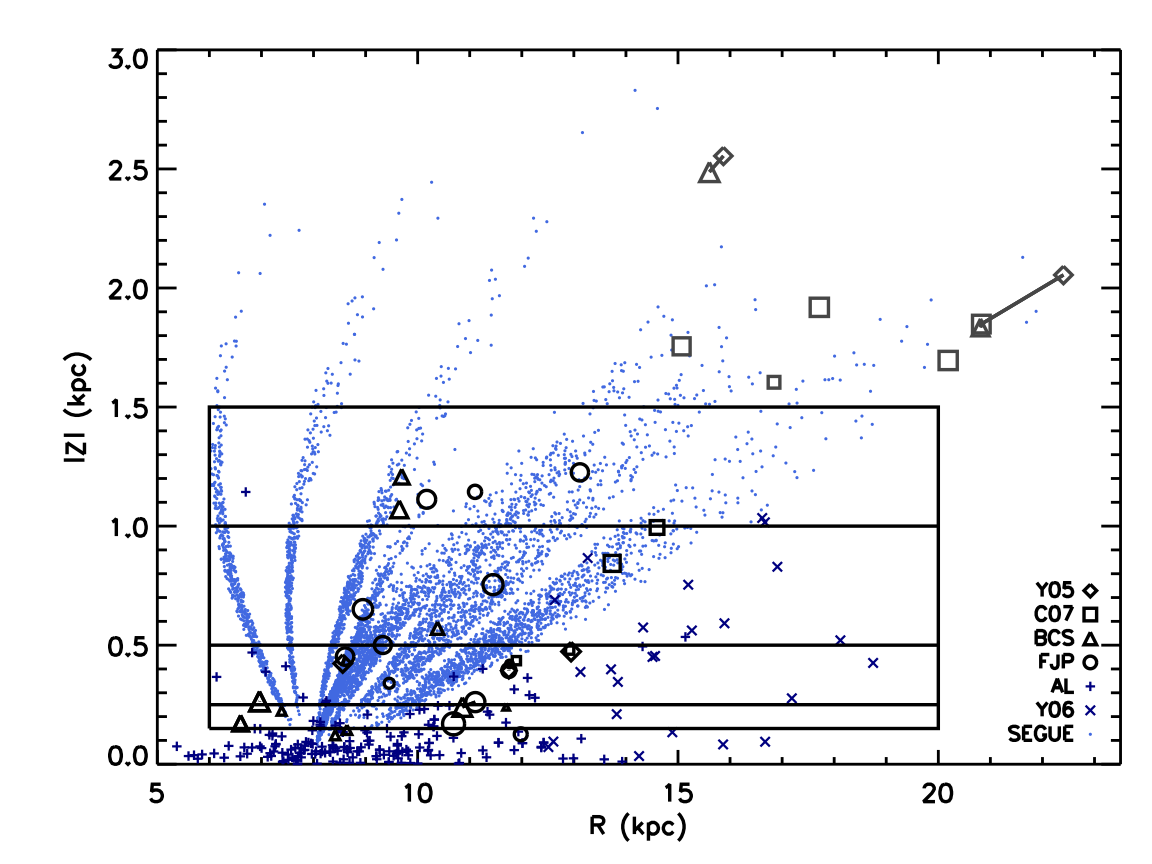}
\caption{Spatial distribution of our sample (blue dots) in Galactic coordinates $R$ and $|Z|$. We divide our sample into four $|Z|$ slices, indicated by the black boxes. Open cluster and Cepheid data are plotted with open symbols and plusses/crosses, respectively (see \S\ref{comparison} for details). Objects from the literature located at $|Z| > 1.5$ kpc are plotted in gray. We note that all of the outer disk clusters from the literature are located at large $|Z|$. Multiple literature values for a single cluster are connected by lines.}
\label{r_z}
\end{figure}

With distances for all of our targets, we obtain the spatial distribution of our sample in Galactic coordinates $R$ and $|Z|$, as shown in Figure~\ref{r_z}.  The SEGUE targets are shown as blue dots. Our radial coverage extends mostly to Galactocentric radii at the solar circle and beyond, though at low $|Z|$ it is confined just to the solar circle. About one-third of our sample is located below the Galactic plane; the $b < 0^{\circ}$ lines of sight cover the entire radial range outside of the solar circle. Thirty-three open clusters and 190 Cepheids analyzed in the literature using high-resolution spectra are also shown (open symbols and crosses, respectively). These data have been used to study the radial metallicity gradient of the disk and will serve as comparison samples. The properties of the open clusters are listed in Table~\ref{clusters}, and the data are described in more detail in \S\ref{comparison}.

\begin{table*}
\caption{Open Clusters with High-Resolution Observations.}
\centering
\scriptsize\begin{tabular}{lrrrrrrrl}
\hline
Cluster & \multicolumn{1}{c}{$l$} & \multicolumn{1}{c}{$b$} & \multicolumn{1}{c}{$d$ (kpc)} & \multicolumn{1}{c}{$R$ (kpc)} & \multicolumn{1}{c}{$Z$ (kpc)} & \multicolumn{1}{c}{[Fe/H]} & \multicolumn{1}{c}{Age (Gyr)} & Reference\\
\hline
Be17 & 176.0 &  -3.6 &  2.7 & 10.7 & -0.17 & -0.2 & 10.1 & \citet{fri05}\\
Be20 & 203.5 & -17.3 &  8.6 & 15.9 & -2.55 & -0.5 &  4.1 & \citet{yon05}\\
Be20 & 203.5 & -17.4 &  8.3 & 15.6 & -2.49 & -0.3 &  6.0 & \citet{ses08}\\
Be22 & 199.9 &  -8.1 &  6.0 & 13.7 & -0.84 & -0.3 &  3.3 & \citet{vil05}\\
Be25 & 226.6 &  -9.7 & 11.4 & 17.7 & -1.92 & -0.2 &  5.0 & \citet{car07}\\
Be29 & 197.9 &   8.0 & 14.8 & 22.4 & +2.05 & -0.5 &  4.3 & \citet{yon05}\\
Be29 & 198.0 &   8.0 & 13.2 & 20.8 & +1.83 & -0.3 &  4.0 & \citet{ses08}\\
Be29 & 198.0 &   8.1 & 13.2 & 20.8 & +1.85 & -0.4 &  4.5 & \citet{car04}\\
Be31 & 206.2 &   5.1 &  5.3 & 12.9 & +0.47 & -0.5 &  5.3 & \citet{yon05}\\
Be32 & 208.0 &   4.4 &  3.4 & 11.1 & +0.26 & -0.3 &  5.9 & \citet{fri10}\\
Be32 & 208.0 &   4.4 &  3.1 & 10.9 & +0.24 & -0.3 &  5.5 & \citet{ses06}\\
Be39 & 223.5 &  10.1 &  4.3 & 11.4 & +0.75 & -0.2 &  7.0 & \citet{fri10}\\
Be73 & 215.3 &  -9.4 &  9.8 & 16.8 & -1.60 & -0.2 &  1.5 & \citet{car07}\\
Be75 & 234.3 & -11.1 &  9.1 & 15.1 & -1.76 & -0.2 &  4.0 & \citet{car07}\\
Cr261 & 301.7 &  -5.5 &  2.8 &  7.0 & -0.26 & +0.1 &  6.0 & \citet{ses08}\\
Cr261 & 301.7 &  -5.5 &  2.8 &  7.0 & -0.27 & \phantom{+}0.0 &  6.0 & \citet{carretta05}\\
M67 & 215.7 &  31.9 &  0.8 &  8.6 & +0.42 & \phantom{+}0.0 &  4.3 & \citet{yon05}\\
M67 & 215.6 &  31.7 &  0.9 &  8.6 & +0.45 & \phantom{+}0.0 &  4.3 & \citet{fri10}\\
Mel66 & 260.5 & -14.2 &  4.4 &  9.6 & -1.07 & -0.3 &  4.0 & \citet{ses08}\\
NGC1193 & 146.8 & -12.2 &  5.8 & 13.1 & -1.23 & -0.2 &  4.2 & \citet{fri10}\\
NGC1817 & 186.1 & -13.1 &  1.5 &  9.5 & -0.34 & -0.1 &  1.1 & \citet{jac09}\\
NGC188 & 122.8 &  22.5 &  1.7 &  8.9 & +0.65 & +0.1 &  6.3 & \citet{fri10}\\
NGC1883 & 163.1 &   6.2 &  3.9 & 11.8 & +0.42 & \phantom{+}0.0 &  0.7 & \citet{jac09}\\
NGC2141 & 198.0 &  -5.8 &  3.9 & 11.8 & -0.39 & \phantom{+}0.0 &  2.4 & \citet{jac09}\\
NGC2141 & 198.1 &  -5.8 &  3.9 & 11.8 & -0.39 & -0.1 &  2.5 & \citet{yon05}\\
NGC2158 & 186.6 &   1.8 &  4.0 & 12.0 & +0.13 & \phantom{+}0.0 &  1.9 & \citet{jac09}\\
NGC2204 & 226.0 & -16.2 &  4.1 & 11.1 & -1.14 & -0.2 &  2.0 & \citet{jac11}\\
NGC2243 & 239.5 & -18.0 &  3.6 & 10.2 & -1.11 & -0.4 &  4.7 & \citet{jac11}\\
NGC2324 & 213.4 &   3.3 &  4.2 & 11.7 & +0.24 & -0.2 &  0.6 & \citet{bra08}\\
NGC2477 & 253.6 &  -5.8 &  1.2 &  8.4 & -0.12 & +0.1 &  1.0 & \citet{bra08}\\
NGC2506 & 230.6 &   9.9 &  3.3 & 10.4 & +0.57 & -0.2 &  1.7 & \citet{carretta04}\\
NGC2660 & 265.9 &  -3.0 &  2.8 &  8.6 & -0.14 & \phantom{+}0.0 &  1.0 & \citet{ses06}\\
NGC3960 & 294.4 &   6.2 &  2.1 &  7.4 & +0.22 & \phantom{+}0.0 &  0.9 & \citet{ses06}\\
NGC6253 & 335.5 &  -6.2 &  1.6 &  6.6 & -0.17 & +0.5 &  3.0 & \citet{carretta07}\\
NGC6253 & 335.5 &  -6.3 &  1.6 &  6.6 & -0.17 & +0.4 &  3.0 & \citet{ses07}\\
NGC6819 &  74.0 &   8.5 &  8.2 &  9.7 & +1.21 & +0.1 &  2.7 & \citet{bra01}\\
NGC7142 & 105.0 &   9.0 &  3.2 &  9.3 & +0.50 & +0.1 &  4.0 & \citet{jac08}\\
Ru4 & 222.0 &  -5.3 &  4.7 & 11.9 & -0.43 & -0.1 &  0.8 & \citet{car07}\\
Ru7 & 225.4 &  -4.6 &  6.0 & 12.9 & -0.48 & -0.3 &  0.8 & \citet{car07}\\
Sau1 & 214.7 &   7.4 & 13.2 & 20.2 & +1.70 & -0.4 &  5.0 & \citet{car04}\\
To2 & 232.0 &  -6.9 &  8.3 & 14.6 & -0.99 & -0.3 &  2.2 & \citet{fri08}\\
\hline
\label{clusters}
\end{tabular}
\end{table*}

We use our sample of MSTO stars to measure the metallicity gradient of the old disk in four different $|Z|$ slices as indicated by the black lines in Figure~\ref{r_z}. Taking thin and thick disk scale heights to be $\sim300$ and 900 pc, respectively, the lower two slices ($0.15 < |Z| < 0.25$ and $0.25 < |Z| < 0.5$ kpc) are dominated by the thin disk. The third slice ($0.5 < |Z| < 1.0$ kpc) is made up of a mix of the thin and thick disks, and the fourth slice ($1.0 < |Z| < 1.5$ kpc) is dominated by the thick disk. Of our sample, 7180 stars fall into these four slices.

Dividing our sample in this way allows a comparison of the radial metallicity gradient of the thin disk to that of the thick disk, as well as to distinguish between \textit{radial} and \textit{vertical} trends. We note that \textit{all} of the distant clusters that have been used to study the behavior of the \textit{radial} metallicity gradient in the outer disk ($R > 15$ kpc) are located at least 1.5 kpc from the Galactic midplane; these are shown in gray in Figure~\ref{r_z}.

\section{Correcting for Selection Biases: Weights}\label{weights}
To use field stars to determine the metallicity distribution in the disk, we must understand how the spectroscopic sample is drawn from the underlying population. As there are many more stars than it was possible for SEGUE to obtain spectra, we must assess whether the spectroscopic sample is truly representative of all of the stars in the disk. It likely is not, because our stars are selected to be the bluer stars in the CMD, making our selection biased against metal-rich and older stars, which have redder colors. The severity of the metallicity bias will depend on the ages of metal-rich stars; the older and more metal-rich they are, the more likely they are to fall out of our sample. To correct for this bias, we employ a weighting scheme to step backward in our sample selection and reconstruct the properties of the underlying parent population. We describe the scheme briefly below; further details are in the Appendix. 

There are three major ways in which the spectroscopic sample is different from the parent population along each line of sight: (1) The photometric objects in regions with the highest extinction were not considered for spectroscopy. (2) Not all candidates for spectroscopy are observed. (3) We observe only MSTO stars using a color cut that is biased against redder metal-rich stars. 

Each star in our sample is given three weights corresponding to the three differences listed above: (1) the \textit{area weight} $W_{\rm A}$, which depends on the coverage of targets on the plane of the sky in each line of sight; (2) the \textit{CMD weight} $W_{\rm CMD}$, which depends on the target's location in the CMD and corrects for the random selection of a subsample of all candidates for spectroscopy that pass the MSTO selection; and (3) the \textit{LF weight} $W_{\rm LF}$, which depends on the target's $T_{\rm eff}$, [Fe/H], and location in the CMD, and corrects for the metallicity bias of the MSTO selection. The total weight $W$ is the product of the three weights $W_{\rm A}$, $W_{\rm CMD}$, and $W_{\rm LF}$. After removing targets with $W=0$, we are left with a sample of 7010 stars. Details about how each of these weights is calculated can be found in the Appendix.

\begin{figure}[!ht]
\epsscale{1.1}
\plotone{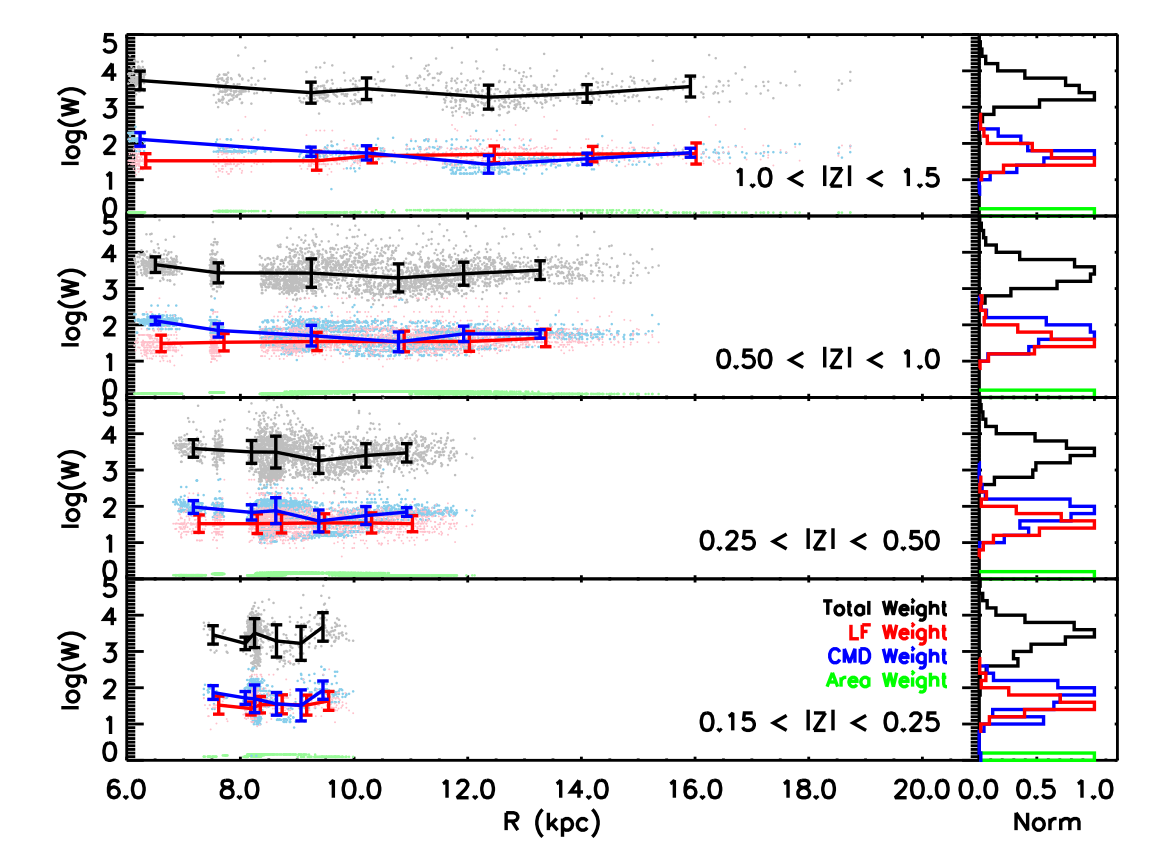}
\caption{Variation in weights as a function of $R$ and $|Z|$. The variation in the total weight $W$ (black) is mostly dependent on the CMD weight $W_{\rm CMD}$ (blue). The LF weight $W_{\rm LF}$ (red) is fairly constant at all locations, with the width of the distribution being $\sim0.5-1.0$ dex. The area weight $W_{\rm A}$ (green) is the smallest contribution and does not vary greatly between different lines of sight. While $W_{\rm CMD}$ shows the most dramatic variation, it is also less uncertain than $W_{\rm LF}$ because it only requires counting objects in CMD bins.}
\label{weights_rz}
\end{figure}

Figure~\ref{weights_rz} shows the distribution of the total weight (black) and each individual weight, as a function of $R$, in four slices of $|Z|$. The panels on the right show the distribution of weights. $W_{\rm A}$ (green) is the smallest contribution and does not vary significantly. $W_{\rm LF}$ (red) is relatively constant as a function of $R$ and $|Z|$, with the distribution having a width of $\sim0.5-1.0$ dex. Since $W_{\rm LF}$ is relatively constant, a systematic error in $W_{\rm LF}$, which could arise from using the wrong luminosity function, will not cause a spurious change in the ratio of metal-poor to metal-rich stars.

$W_{\rm CMD}$ (blue) shows the most variation because it normalizes for the fact that there are more stars in the inner disk than the outer disk. Although the variations in $W_{\rm CMD}$ are large, this weight is less uncertain than $W_{\rm LF}$, as it only requires counting objects in each bin. Figure~\ref{weights_rz} shows that the change in the total weight $W$ mostly follows the change in $W_{\rm CMD}$.

\section{Results: Radial Metallicity Gradients}\label{results}
The total weights are applied to each target, allowing us to use our sample of MSTO stars to estimate the properties of the underlying parent population. We divide our sample into four slices of $|Z|$: (1) $0.15 < |Z| < 0.25$, (2) $0.25 < |Z| < 0.5$, (3) $0.5 < |Z| < 1.0$, and (4) $1.0 < |Z| < 1.5$ kpc. Within each $|Z|$ slice, we fit a linear gradient to the data, weighting each target by the total weight determined using the scheme described in \S\ref{weights}.

\begin{figure*}[!ht]
\epsscale{1.1}
\plotone{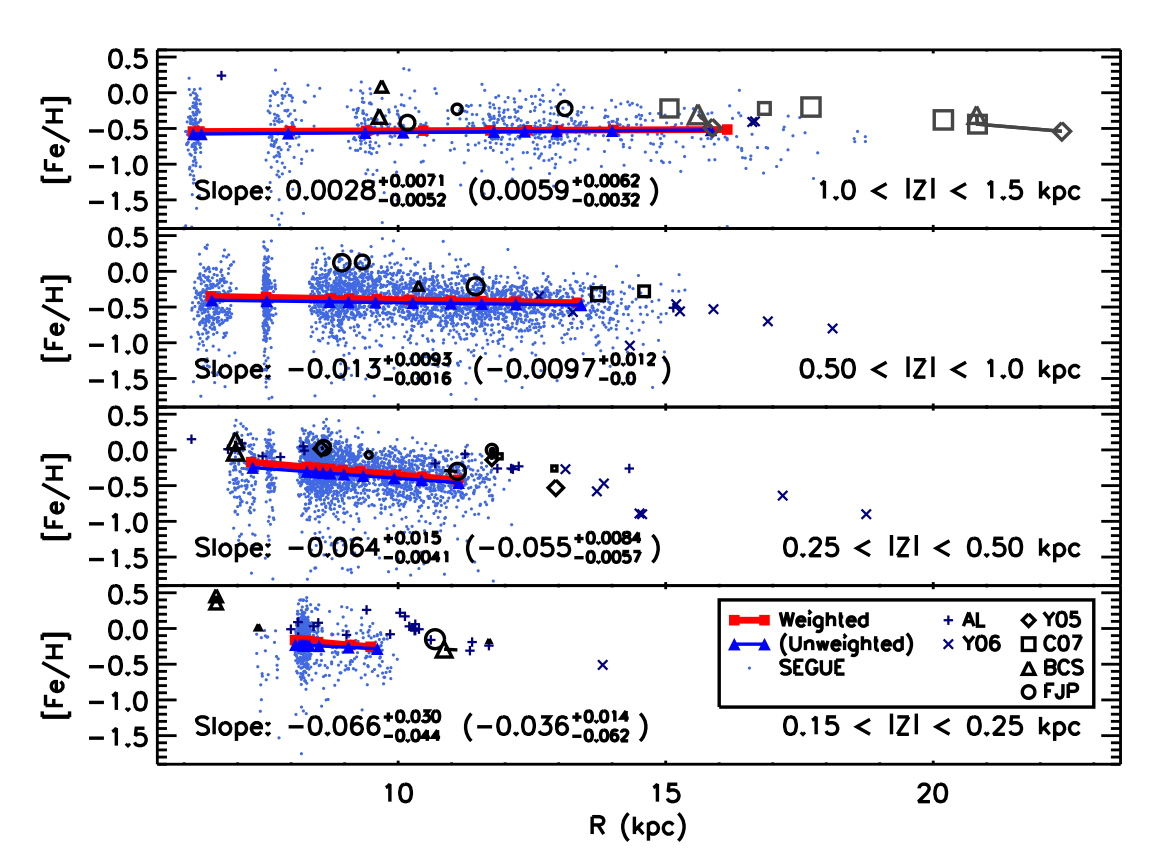}
\caption{Metallicity [Fe/H] vs. Galactocentric radius $R$ in four $|Z|$ slices. Light blue points indicate the SEGUE data. The weighted median metallicity and the derived linear fit are shown as red squares, with the numerical values in the bottom left of each panel. The blue triangles and values in parentheses show the results we would have obtained if no corrections for known selection effects had been applied. The spacing of the symbols indicates the radial distribution of the targets. Open symbols and plusses/crosses are open clusters and Cepheids from the literature (see \S\ref{comparison} for details). The sizes of the open cluster symbols indicate their ages (smaller symbols for younger clusters). At low $|Z|$ ($<0.5$ kpc) our derived gradient is consistent with published values. At high $|Z|$ ($>0.5$ kpc) the constant [Fe/H] is consistent with the cluster metallicities reported by \citet{yon05} in the outer disk.}
\label{r_feh}
\end{figure*}

Figure~\ref{r_feh} shows the radial metallicity gradients for all four $|Z|$ slices for both the unweighted and weighted cases (blue triangles and red squares, respectively). The weighted slopes are written in the bottom left corner of each panel, with the unweighted values in parentheses. The quoted errors are derived from the Monte Carlo simulations described in \S\ref{gradient_errors} and include only the random errors from uncertainties in the stellar parameters. See \S\ref{gradient_errors} for a discussion of the systematic errors. The large symbols and navy blue plusses/crosses show the positions and metallicities for open clusters and Cepheids from the literature (see \S\ref{comparison}).

In the low $|Z|$ slices ($|Z| < 0.5$ kpc), we obtain values that are consistent with the $-0.06$ dex kpc$^{-1}$ determined by \citet{fri02} for open clusters and \citet{luc06} for cepheids.  Our major result is that in the high $|Z|$ slices ($|Z| > 0.5$ kpc), the slope is flat for the entire radial range $6 < R < 16$ kpc. The constant median [Fe/H] in the highest slice ($|Z| > 1.0$ kpc) is consistent with the metallicities reported by \citet{yon05} and others at large $R$. 

\begin{figure}[!ht]
\epsscale{1.0}
\plotone{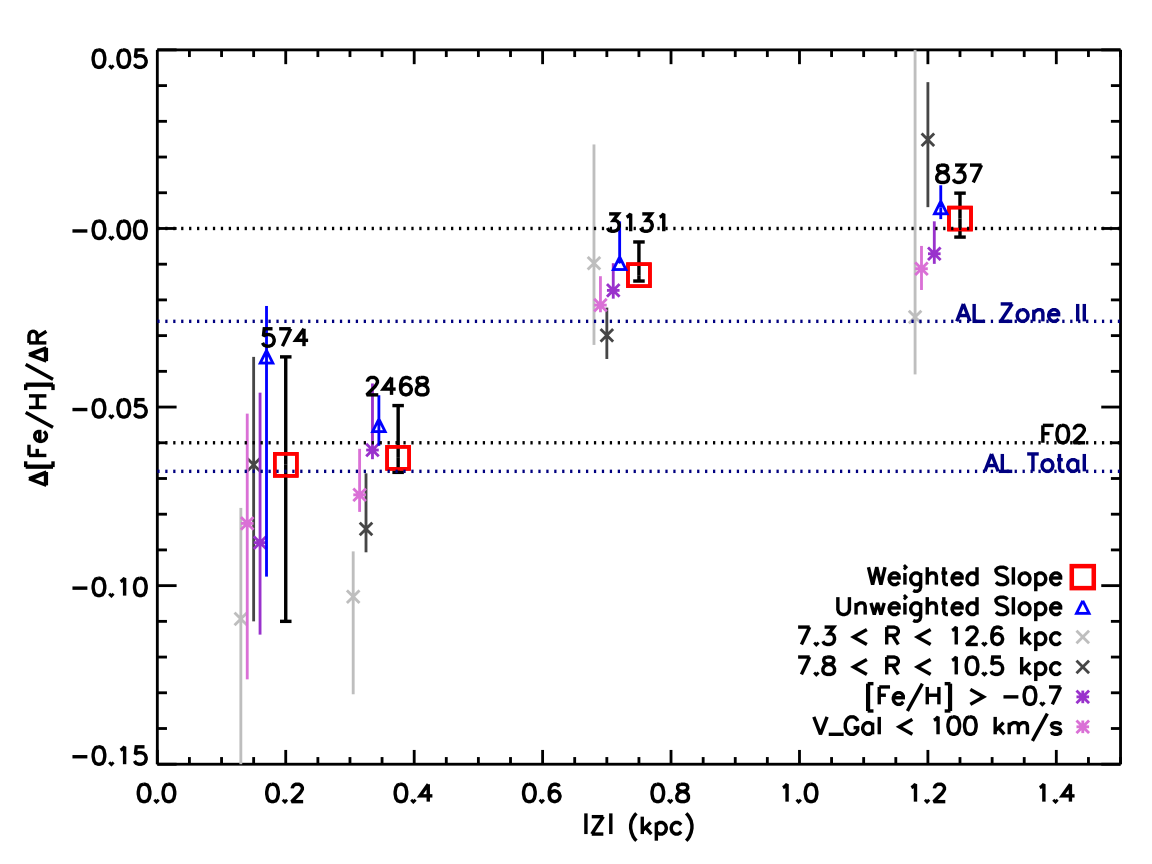}
\caption{Measured gradient $\Delta$[Fe/H]/$\Delta R$ vs. vertical height $|Z|$. The horizontal black dotted lines indicate a flat gradient and the gradient measurement published by \citet{fri02} using open clusters. The horizontal blue dotted lines indicate the gradient measurements published by \citet{luc06} using Cepheids for the ranges $6.6 < R < 10.6$ kpc (zone II) and $4.0 < R < 14.6$ kpc (their total sample). The red squares and blue triangles show the measured gradients (weighted and unweighted, respectively) as a function of height above the Galactic plane presented in this work. The slope becomes flat at high $|Z|$ using both the weighted and unweighted results. The number of objects in each $|Z|$ slice is indicated above the symbols. The gray symbols show the results obtained for restricted ranges in $R$. The purple symbols show the results obtained when probable halo contaminants are removed (see \S\ref{halo}). In each case, the trend of the gradient becoming more shallow with distance from the plane is still observed, which indicates that our result does not arise due to the increase in the radial range or contamination from the halo.}
\label{slopes}
\end{figure}

Our results are summarized in Figure~\ref{slopes}, which shows the measured slopes as a function of vertical height $|Z|$. The horizontal dotted lines show the values in the literature published by \citet{fri02} and \citet{luc06} for open clusters and cepheids, respectively, which are consistent with the values we obtained for the low $|Z|$ slices. Again, the red squares show the values obtained after we apply our weights, and the blue triangles show the values obtained using the unweighted data without any corrections for the selection function. The observed trend---flattening slope with increasing $|Z|$---is seen in both the unweighted and weighted cases. 

To test the robustness of this result, we re-measured the gradient for various subsamples of our data. The purple symbols in Figure~\ref{slopes} show the results if we exclude likely halo stars (see \S\ref{halo}). The gray symbols show the values obtained for restricted ranges in $R$ ($7.3 < R < 12.6$ kpc and $7.8 < R < 10.5$ kpc). These measurements were done to test that the flattening of the radial gradient is truly a trend in $|Z|$, and not just a result of the wider range in $R$ that is observed in the high $|Z|$ bins. The two restricted ranges in $R$ correspond to the extent of observations in the two lower $|Z|$ bins. Figure~\ref{slopes} shows that gradient becomes flat with or without weighting and regardless of the range in $R$ observed.

\section{Errors}\label{errors}
\subsection{Metallicity Bias}\label{metalbias}

\begin{figure*}[!ht]
\epsscale{0.55}
\plotone{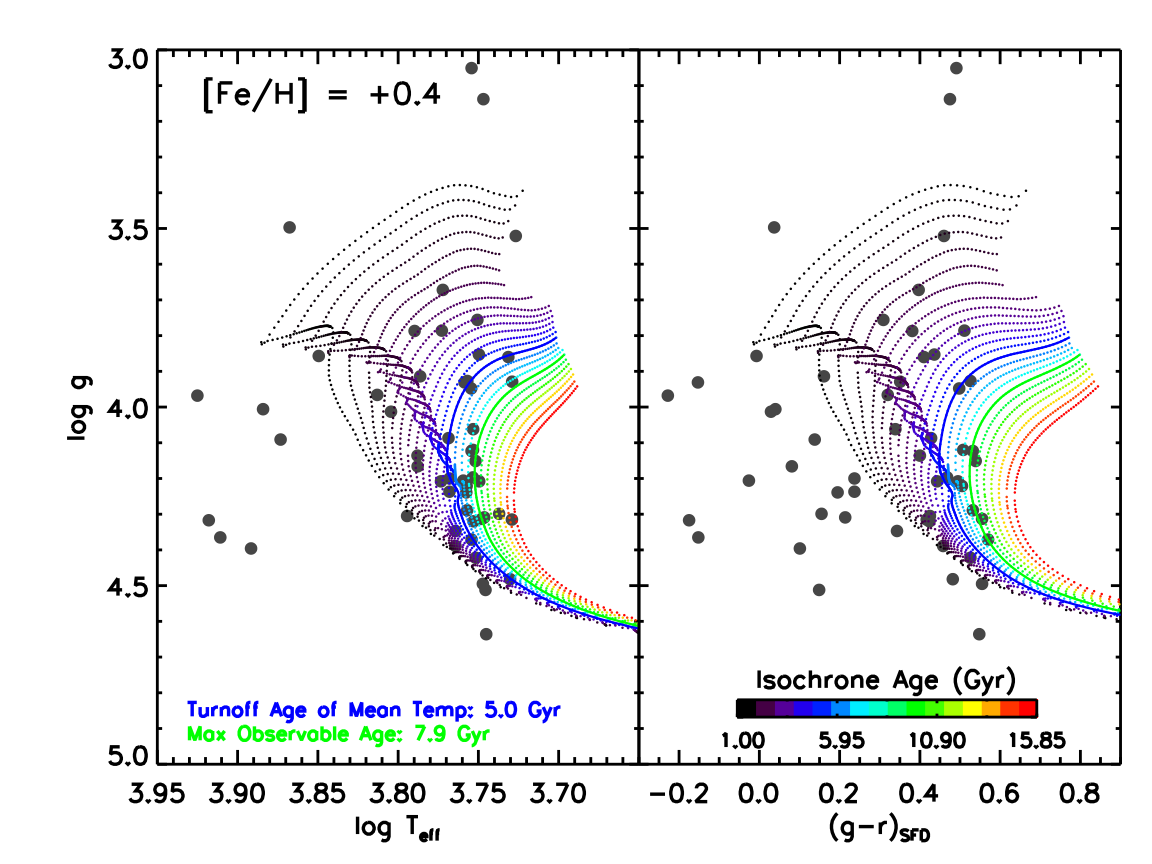}
\plotone{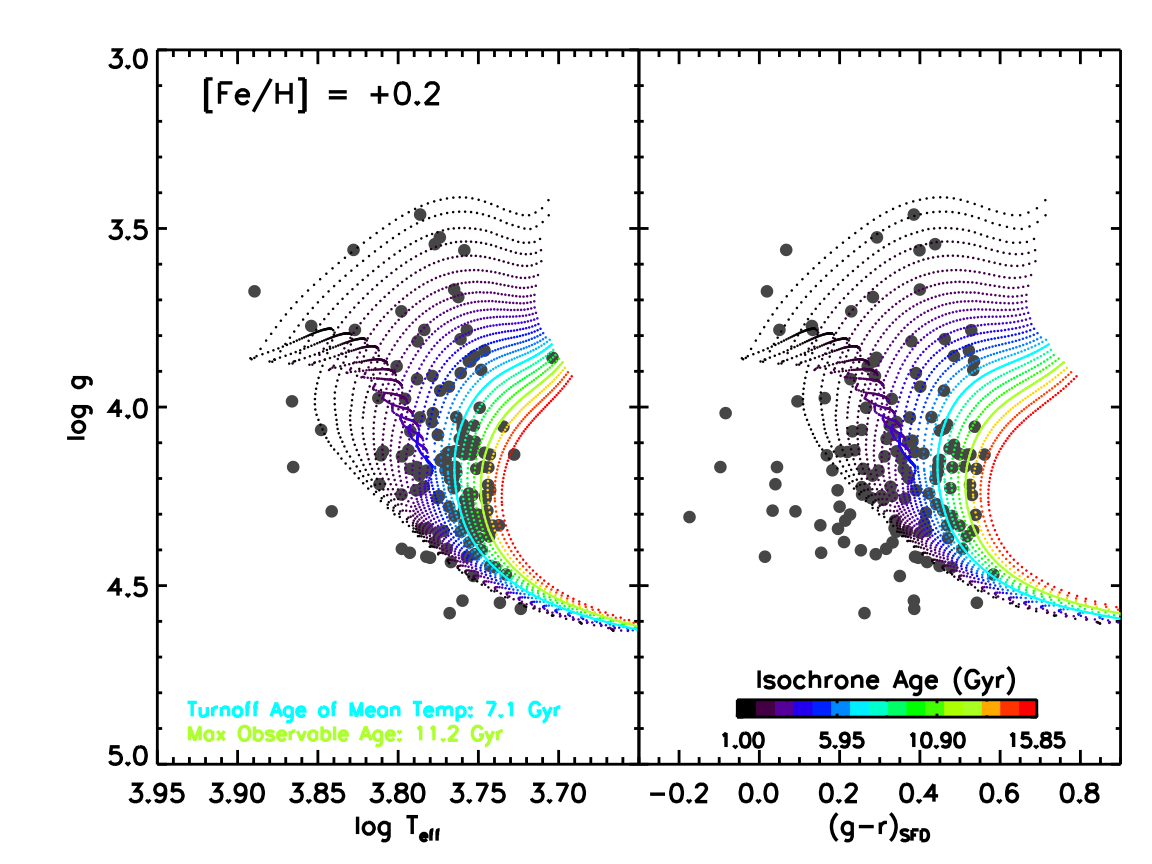}
\caption{Maximum observable ages for metal-rich stars. We compare the SEGUE data (gray circles) with theoretical isochrones (color, \citealt{an09}) to estimate the oldest main sequence turnoff stars that would fall in our sample. The TAMT (see Table~\ref{meanages}) and maximum observable age are shown as solid lines. For [Fe/H]$=+0.4$ (left panel), the coolest/reddest stars may be as old as $\sim8$ Gyr; for [Fe/H]$=+0.2$ (right panel), the coolest/reddest stars that fall in our sample may be as old as $\sim11$ Gyr. Based on the measured ages of nearby stars \citep{ben05,cas11}, we do not expect there to be a significant population of old, metal-rich stars which would be excluded from our sample.}
\label{maxage}
\end{figure*}

As described in \S\ref{targets}, our sample of MSTO stars is chosen by making a cut in $(g-r)_{\rm SFD}$, which likely biases it against redder, more metal-rich stars. Theoretical isochrones give us an idea of the significance of this bias. Figure~\ref{maxage} shows the temperature and $(g-r)_{\rm SFD}$ distributions of metal-rich stars in our sample ([Fe/H] $\geq+0.2$, gray circles) against the theoretical isochrones of \citet[color]{an09}. We estimate that the maximum observable ages of stars at [Fe/H] $=+0.4$ and $+0.2$ are $\sim8$ and 11 Gyr, respectively. At each metallicity, this corresponds to the turnoff age of the coolest/reddest stars that pass the selection criteria for our sample. The TAMT and maximum observable age are shown as solid lines. Based on the measured ages of nearby stars \citep{ben05,cas11}, very few stars with metallicities [Fe/H] $> +0.2$ are older than 8 Gyr. We therefore expect that metal-rich stars are well represented in our sample. For stars with metallicities [Fe/H] $\leq$ 0.0, there is no significant bias against stars of any age.

The arguments presented above suggest that the metallicity bias resulting from our color selection does not eliminate a significant fraction of the disk population from our sample.  However, the color selection also causes metal-rich stars to be under-represented in our data. We use mock catalogs from the models of \citet{sch09a} to quantify how well we correct for this effect and recover the true metallicity distribution in our sample. We show below (\S\ref{gradient_errors}) that we are able to measure the true gradients within the errors and reproduce the trends as a function $|Z|$ when we apply the weighting scheme described in \S\ref{weights} and the Appendix.

\subsection{Halo Contamination}\label{halo}
Since we are interested in the metallicity distribution of the Galactic disk, we must assess whether our sample may be contaminated by halo objects. Our sample does not reach $R < 5$ kpc, where the bulge is expected to be a significant population. One way to quantify the amount of contamination is by examining different multi-component models for the Galaxy, in particular, those of \citet{jur08} and \citet{dej10}. For the lines of sight in our sample, these two models predict total halo contaminations of $\sim2\%$ and $\sim0.8\%$, respectively. In the most distant $|Z|$ bin, the predicted contaminations increase to $\sim11\%$ and $\sim5.6\%$, respectively. The difference in the predictions can be almost entirely attributed to the discrepant local densities for the halo that were derived (0.51\% by \citealt{jur08} and 0.17\% by \citealt{dej10}).

Both models, however, predict that the anti-center lines of sight, aimed toward the outer disk, will be more contaminated by halo stars. This is especially true in the \citet{jur08} model, which predicts that the two lines of sight that reach the largest values of $R$ will have $17-18\%$ contamination (compared to $9-13\%$ in the other lines of sight). If there is indeed more halo contamination in the outer regions, this should push our measured gradient to be steeper, since there will be more metal-poor halo stars at large $R$ compared to the inner disk. However, we see the opposite effect, that the gradient is flatter in the highest $|Z|$ bins.

Looking at the data, we do not see evidence for significant halo contamination. Halo stars, which have large velocities and are more metal poor than disk stars, can generally be identified using their kinematics or chemistry. To assess the effect that such stars may have on our results, we recalculate the gradient after applying two different cuts to remove potential halo stars from the data. (1) A metallicity cut that removes all stars with [Fe/H] $<-0.7$, and (2) a kinematic cut that removes all stars with $V_{\rm Gal} < 100$ km s$^{-1}$ to remove stars with the largest velocity offset relative to the projection of the local standard of rest, where $V_{\rm Gal} = V_R+220\cdot{\rm cos}b\cdot{\rm sin}l$ and $V_R$ is the line-of-sight velocity measured from the SEGUE spectra. We only remove stars with $V_{\rm Gal} < 100$ km s$^{-1}$ in lines of sight with $50 < l < 130^{\circ}$. We do not include the lines of sight directed toward the Galactic anti-center because the local standard of rest is tangent to those directions and the projection does not give a meaningful velocity. The two cuts remove $\sim700$ and $\sim100$ stars, respectively. The gradients we measure using each of these cuts are not significantly different from our main results, as shown by the purple symbols in Figure~\ref{slopes}.

\subsection{Mock Catalog Analysis}
To quantify the errors in our analysis, we utilize mock observations generated from the models of \citet{sch09a}. By using a model where we know the true stellar parameters, distances, and metallicity distributions of the targets, we can test whether we are able to reliably reproduce the ground truth after applying our observational selection to the mock catalogs and performing the same analysis procedures. As the purpose is merely to assess how accurately we can measure quantities such as distances and metallicity gradients, this way of testing our methods is not dependent on having a \textit{correct} model for the Galaxy. We do, however, need a model that can reproduce some basic properties of the observed stellar populations. \citet{sch09a} have shown that their model provides a good match to the properties of stars in the solar neighborhood as observed by the Geneva-Copenhagen Survey \citep{nor04}. 

\begin{figure*}[!ht]
\epsscale{0.55}
\plotone{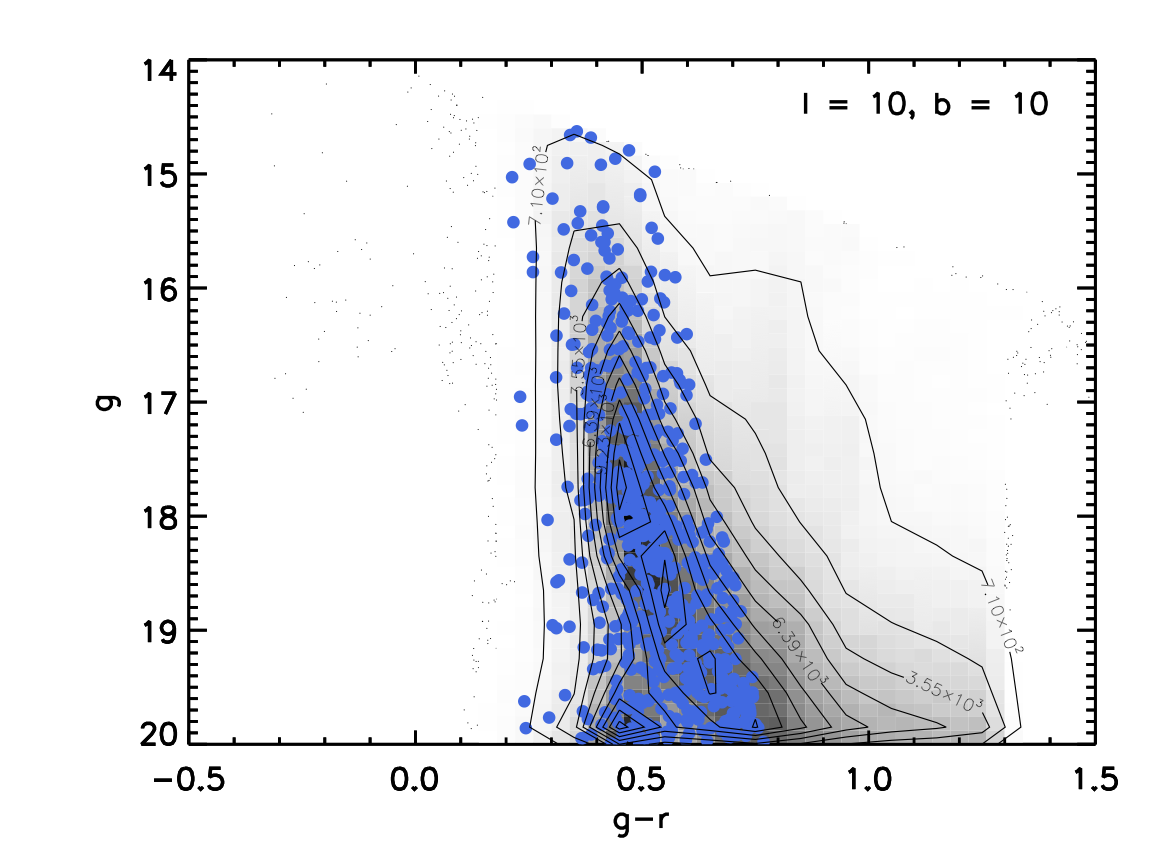}
\plotone{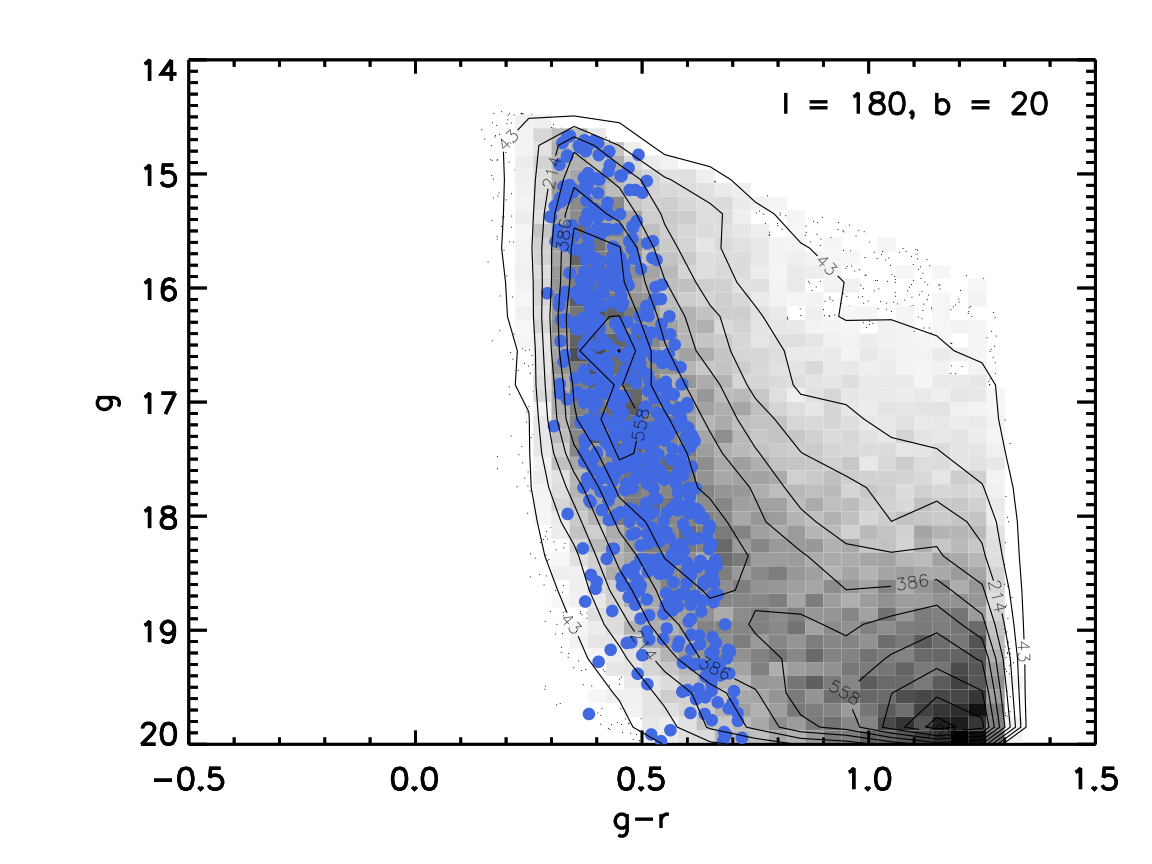}
\caption{Mock catalog MSTO selection. This figure is analogous to Figure~\ref{cmd_selection}, but using simulated observations from \citet{sch09a}. The full sample of stars (grayscale and contours) is put through the same sample selection as the real data to yield an analogous MSTO sample (blue circles). The similarity between the CMDs of the real and mock data shows that we are using a reasonable model of the Galaxy to test our errors.}
\label{mock_cmd}
\end{figure*}

We use a mock catalog, provided by R. Sch{\"o}nrich, of 6,701,170 stars with $ugriz$ colors, stellar parameters, and distances along ten lines of sight. We repeat the same sample selection described in \S\ref{targets} on the mock catalog to replicate the effects of using MSTO stars as tracers. Figure~\ref{mock_cmd} is analogous to Figure~\ref{cmd_selection} and shows the results of the random subsampling for two lines of sight. Though the model was not tuned to look like the SEGUE data, the model CMDs are good matches to the observations.

In our analysis, we use all 111,640 objects that fulfill the MSTO selection to estimate errors on the distance (\S\ref{distanceerrors}). We draw many random subsamples of 6500 MSTO stars to estimate the systematic and random errors on our gradient measurement (\S\ref{gradient_errors}), to simulate the effect of having a limited number of spectroscopic targets in each line of sight. We repeat the analysis using different random subsamples to assess how much the results change based on which particular targets are chosen for spectroscopy.

\subsubsection{Errors in Distance Estimates}\label{distanceerrors}
\textbf{Systematic Errors:} The main source of systematic error in the distance arises from assumptions we must make when choosing isochrones to estimate the luminosities of our stars. One source of error is the $\alpha$-enhancement; \citet{an09} adopt an $\alpha$-enhancement scheme where each value of [Fe/H] has an assumed value of [$\alpha$/Fe] (see Table 2). We do not expect the discrepancy between the target and isochrone [$\alpha$/Fe] to have a large effect on the distance estimate. The [$\alpha$/Fe] of stars in our sample \citep{lee11a} are generally within 0.2 dex of the values assigned to the \citet{an09} isochrones. Using the [Fe/H]-[$\alpha$/Fe] grid of \citet{dot08}\footnote{\footnotesize{http://stellar.dartmouth.edu/$\sim$models/grid.html}}, we estimate that a 0.2 dex change in [$\alpha$/Fe], at worst, leads to a $\sim10\%$ change in the distance for a star on the zero-age main sequence.

\begin{figure}[!ht]
\epsscale{1.1}
\plotone{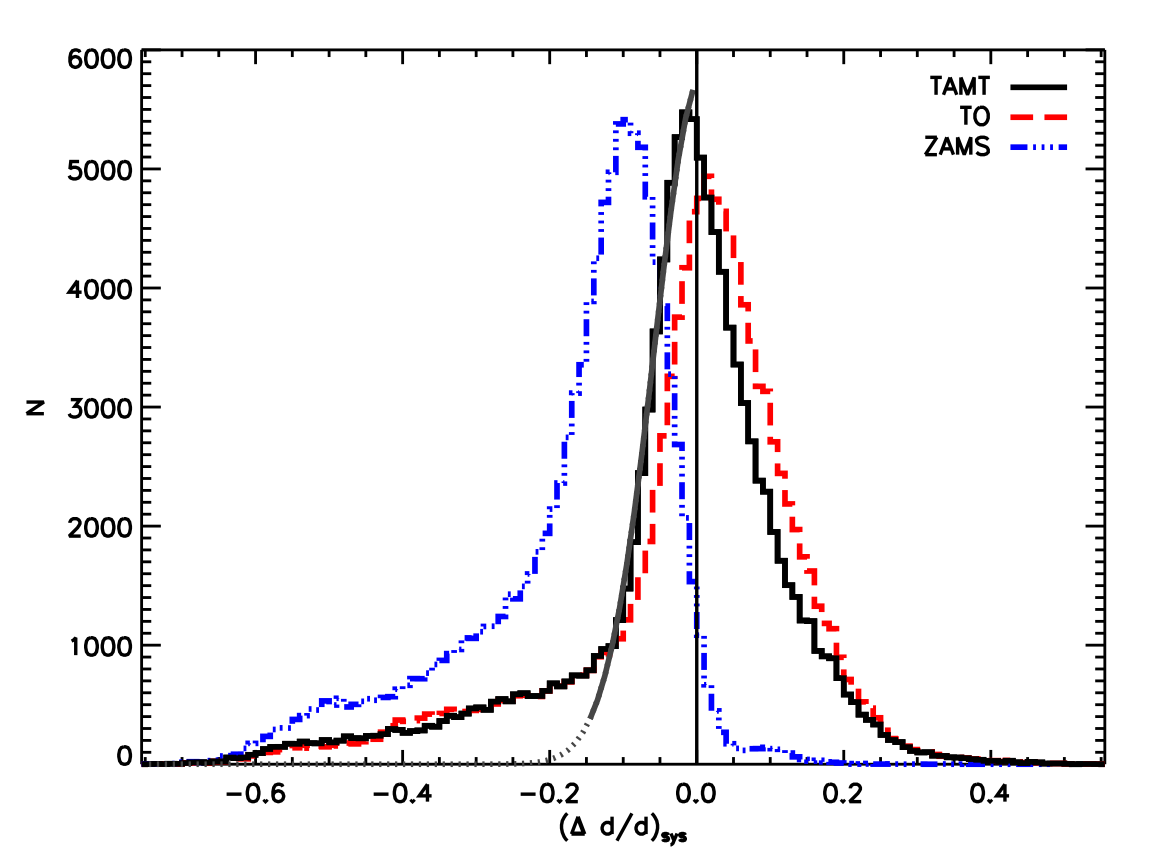}
\caption{The fractional error distributions of solar metallicity stars ($-0.05 < $ [Fe/H] $ < 0.05$) for three different age assumptions, as described in the text (see also Figure~\ref{isochrone_ages}). The TAMT age assumption gives the best agreement between the calculated and true model distances. Based on the width of the distribution, we estimate the systematic error on our distance estimates to be $\sim 10\%$. The vertical line indicates where the fractional error is zero. The gray line is a Gaussian fit to the TAMT distribution where $-0.15 < (\Delta d/d)_{\rm sys} < 0.0$, which we use to estimate the fraction of subgiants in our sample ($\sim15\%$).}
\label{sp_vs_to}
\end{figure}

For our sample, a larger source of uncertainty is the ages we assign to our targets; the slope of the main sequence becomes very steep at the turnoff, making the predicted distance a strong function of the chosen age. We compare the distance estimates obtained using three different age assumptions, which are shown schematically for stars at solar metallicity ($-0.05<$ [Fe/H] $<0.05$) in Figure~\ref{isochrone_ages}: 

\begin{enumerate}
\item Turnoff Age of the Mean Temperature (TAMT): We use one isochrone for most of the stars in the metallicity bin---this is determined by finding the mean of the temperature distribution. Targets hotter than the main sequence turnoff are assigned to the oldest possible isochrone. This was the scheme used throughout our analysis (described in \S\ref{distances}) and illustrated in the second panel of Figure~\ref{isochrone_ages}.
\item Turnoff (TO): We use the oldest possible isochrone consistent with their measured temperature for all targets---this assumes that every star is located at the turnoff. In comparison to the TAMT assumption, this approach changes the distances for stars cooler than the mean temperature of the sample, as shown in the third panel of Figure~\ref{isochrone_ages}.
\item Zero-Age Main Sequence (ZAMS): We use the youngest possible isochrone for all targets---this assumes that most targets are located on the zero-age main sequence, as shown in the fourth panel of Figure~\ref{isochrone_ages}.
\end{enumerate}

For this analysis, we compare our calculated distance $d$ to the true distance from the model $d_{\rm model}$ for 111,640 mock targets at solar metallicity ($-0.05<$ [Fe/H] $<0.05$), where theoretical isochrones from different groups show the best agreement. Any disagreement between the isochrones used to generate the mock catalog and the isochrones used to calculate the distances will introduce an additional systematic error. The distributions of the systematic fractional error $(\Delta d/d)_{\rm sys}$ for all three age assumptions are shown in Figure~\ref{sp_vs_to}, where the systematic fractional error is given by:
\begin{equation}
\left(\frac{\Delta d}{d}\right)_{\rm sys} = \frac{d-d_{\rm model}}{d_{\rm model}}
\end{equation}

\noindent Based on the offsets of the distributions, the TO (red dashed line) and ZAMS (blue dash-dotted line) assumptions overestimate and underestimate the distances, respectively, though the ZAMS assumption does much worse. The TAMT (black solid line) assumption gives the best agreement between the calculated and model distances, with an error of $\sim 10\%$, based on the width of the $(\Delta d/d)_{\rm sys}$ distribution. Mock targets at other metallicities exhibit the same behavior, with the TAMT assumption giving the best distances with a mean error of $\sim 10\%$. This error includes the discrepancy between the isochrone [$\alpha$/Fe] and that of individual stars, as the model stars have a range of [$\alpha$/Fe] that do not exactly match the sequence of the \citet{an09} isochrones.

Using the $(\Delta d/d)_{\rm sys}$ distribution, we can also estimate the number of subgiants in our sample by investigating the long tail of negative fractional errors. The gray line in Figure~\ref{sp_vs_to} shows a Gaussian fit to the TAMT histogram where $-0.15 < (\Delta d/d)_{\rm sys} < 0.0$---this range was chosen by eye to reflect the range where subgiants were not contributing to the distribution. By examining the discrepancy between the Gaussian fit and the long tail of negative $(\Delta d/d)_{\rm sys}$ values, we estimate the contamination by subgiants to be $\sim15\%$; we obtain the same result if we take subgiants to be all stars with $(\Delta d/d)_{\rm sys}\leq-0.15$. We derive similar values for other metallicities as well. 

We can also assess how an error in the value of the TAMT affects the distances that we obtain in our calculation. If we view the TO assumption as the TAMT assumption with an incorrect TAMT---one that is too old by 0.2 dex---we can estimate how much the distances change as a result. Comparing the TAMT and TO results in Figure~\ref{sp_vs_to}, the peak $(\Delta d/d)_{\rm sys}$ values of the two assumptions are within 5$\%$, which suggests that any error in the TAMT is smaller than the error from the age assumption scheme used.

\textbf{Random Errors:} The random error in the distances is dominated by uncertainties in the stellar parameters derived by the SSPP; errors in the magnitudes are trivial in comparison. We examine how the parameter errors propagate through our analysis by generating 500 Monte Carlo realizations of the mock catalog with slightly perturbed values of the stellar parameters and repeating the same distance measurements each time. For each realization we assign a Gaussian distribution of perturbations with widths of 200 K and 0.3 dex for $T_{\rm eff}$ and [Fe/H], respectively. These values are motivated by the errors estimated by \citet{lee08a}. The resulting random fractional error $(\Delta d/d)_{\rm rand}$, given by:
\begin{equation}
\left(\frac{\Delta d}{d}\right)_{\rm rand} = \frac{d_{\rm perturbed}-d}{d}
\end{equation}

\noindent where $d_{\rm perturbed}$ is the distance calculated with the new perturbed values of the stellar parameters. From the distribution of $(\Delta d/d)_{\rm rand}$, we estimate that the uncertainty in the SSPP parameters translates to a $\sim 15-20\%$ error in the distance. These values correspond to the 68\% confidence levels derived from the width of the distribution.

We can now calculate the total error in the distance, which combines the systematic and random errors from the age assumption and the stellar parameters, respectively. The result is the total fractional error $(\Delta d/d)_{\rm total}$, given by:
\begin{equation}
\left(\frac{\Delta d}{d}\right)_{\rm total} = \frac{d_{\rm perturbed}-d_{\rm model}}{d_{\rm model}}
\end{equation}

\noindent We estimate that including both the systematic and random errors, we have a total distance error of $\sim 20-25\%$, with a larger contribution from the uncertainty in the stellar parameters determined by the SSPP. This total error is comparable to the size of the errors in the distances estimated by \citet{gil95} for their sample of MSTO stars.

\subsubsection{Errors in the Metallicity Gradient Measurement}\label{gradient_errors}
\begin{figure}[!ht]
\epsscale{1.0}
\plotone{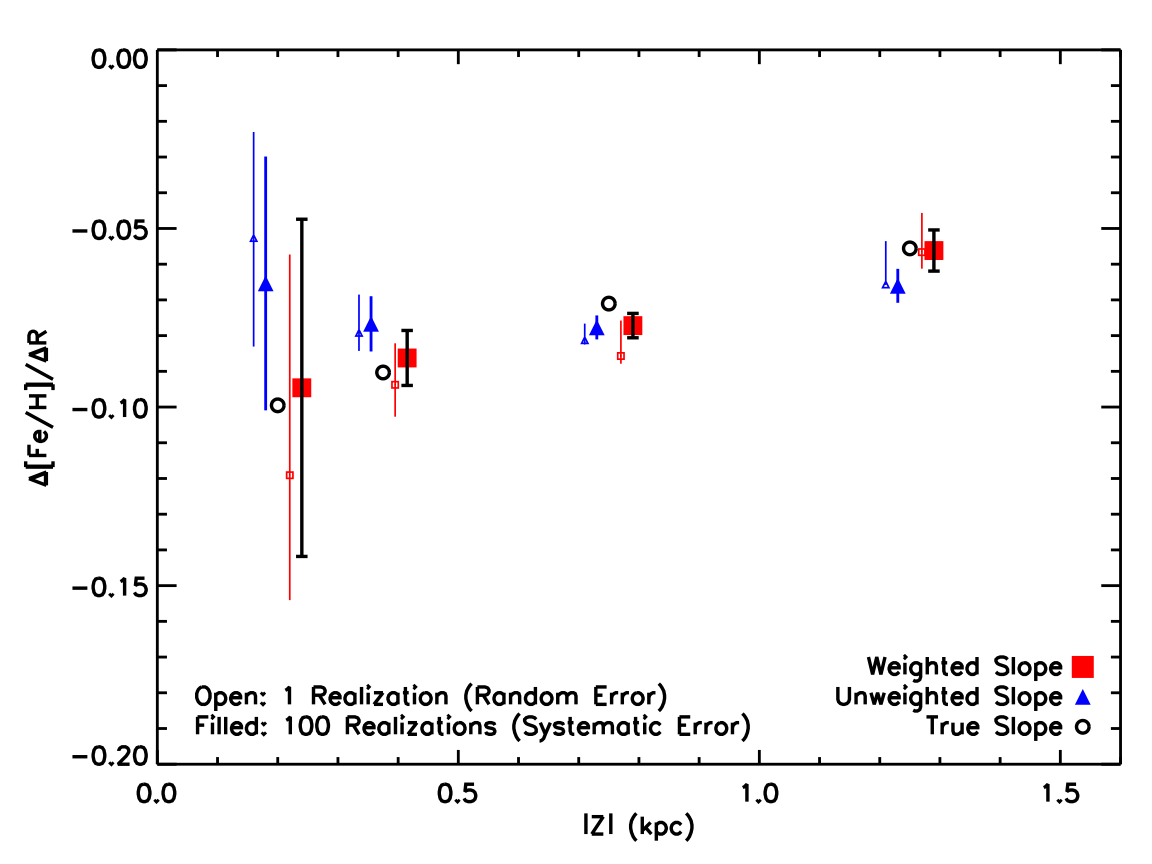}
\caption{Measured gradient $\Delta$[Fe/H]/$\Delta R$ vs. vertical height $|Z|$ using mock samples. This figure is analogous to Figure~\ref{slopes}, except the samples used are drawn from mock catalogs of the galaxy model of \citet{sch09a}. The mean and standard deviation for 100 such samples of MSTO stars are shown as filled symbols. The open symbols show the measured gradients for one particular random sample, with the error bars showing the random errors derived from Monte Carlo realizations using perturbed stellar parameters. The black circles show the true gradient, which is measured using all of the targets in the catalog. The true gradient generally falls within the errors presented for the weighted gradient (red squares) measured using the MSTO samples, which indicates that our method of measuring the gradient gives a reliable result. More importantly, we are able to reproduce the flattening trend seen in the true gradient. This is not true for the unweighted gradient measurements (blue triangles). Note that the vertical scale is different from that of Figure~\ref{slopes}.}
\label{mock_slopes}
\end{figure}

\textbf{Systematic Errors:} Using the same mock catalog, we also assess how well we measure the radial metallicity gradients in the disk. Figure~\ref{mock_slopes} is the same as Figure~\ref{slopes} but shows the gradient results for the mock catalogs in four $|Z|$ slices. The true gradients, measured using the entire mock catalog, are shown as black circles. We draw 100 random samples (6500 MSTO stars each) from the mock catalog using the same selection criteria as the real data (\S\ref{targets}) and follow the same analysis procedures, including accounting for the weights (\S\ref{weights} and the Appendix). 

The filled symbols in Figure~\ref{mock_slopes} show the mean gradient of the 100 realizations; the error bars indicate the standard deviation of the distribution. The smaller open symbols show the gradient measurements for one particular realization, with the error bar indicating the random error due to the uncertainty in the stellar parameters (see below). 

The true gradients (open black circles) are generally within the errors of the weighted measurements (filled red squares), indicating that our measurements are reliable. More importantly, the weighted measurements show the same flattening trend with $|Z|$ as the true gradient. This is not seen in the unweighted gradient measurements (filled blue triangles). While the metallicity gradient in the model is steep compared to recent observations (e.g., \citealt{luc11}), our mock catalog analysis demonstrates that the weighting procedure is necessary to reproduce the ßattening trend. 

\textbf{Random Errors:} The error bars on the open symbols represent the random errors, which are derived by running our full analysis on one sample of 6500 MSTO stars from the mock catalog, where we generate 500 Monte Carlo realizations of the mock data with perturbed stellar parameters. The error bars indicate the 68$\%$ confidence levels derived from the width of the distribution of slopes we obtain in the 500 realizations. This is the same procedure followed to determine the random distance errors in \S\ref{distanceerrors}. The random errors are comparable to the systematic errors on the gradient, although in some cases they are larger. The large error bars for the lowest $|Z|$ bin is due to a combination of the smaller number of stars and the smaller range in $R$.

\section{Discussion}\label{discussion}
\subsection{Comparison with Previous Studies}
The median metallicity we find at $|Z|>1$ kpc  ([Fe/H] $\sim-0.5$) is consistent with the value published by \citet{gil95} for their stars located 1.5 kpc above the plane; the median $|Z|$ for our sample above 1 kpc is 1.24 kpc. Their sample also consisted of an in-situ sample of F/G dwarfs, allowing for a direct comparison. The G dwarf sample of \citet{lee11b}, also drawn from SEGUE, has median values of [Fe/H] within 0.1 dex of that of our sample as a function of $|Z|$.

Our results are also consistent with the lack of a radial metallicity gradient found by both \citet{all06}, using spectra of 12,483 F/G stars from the SDSS, and \citet{jur08}, using photometric metallicities for millions of stars in the SDSS, but both studies examine a much higher $|Z|$ sample with a limited $R$ range at the vertical heights that our sample covers. The present study is complementary to these two studies in that we observe lower Galactic latitudes and we are able to make a direct comparison of the thin and thick disks using the same homogeneous sample. Our results are based on a sample that has been carefully corrected for selection effects and we use an improved version of the SSPP that is more accurate for metal-rich stars, which was not available for these previous analyses.

In their analysis of thick disk stars in the solar neighborhood, \citet{ben03,ben05} find that abundance trends using O, Mg, and Fe do not vary as a function of $R$ or $|Z_{\rm max}|$ in the region $5 < R < 7$ kpc, $0 < |Z_{\rm max}| < 1.1$ kpc, where $|Z_{\rm max}|$ is the maximum vertical height reached by the calculated stellar orbit. These detailed abundances, together with our finding of a flat metallicity gradient at large $|Z|$, are suggestive of a chemically homogeneous thick disk, although more observational data are needed to confirm this.

\subsection{Comparison with Open Clusters and Cepheids}\label{comparison}
In Figures~\ref{r_z} and \ref{r_feh} we show open cluster and Cepheid data from the literature. These two classes of objects have been studied extensively with high-resolution spectroscopy and also span a large range in age, giving them the power to probe the chemistry of the interstellar medium at different times. Cepheids typically have lifetimes on the order of $\sim100$ Myr and trace the present day metallicity distribution. Open clusters can have ages anywhere from less than a Gyr to greater than 10 Gyr and have been used to examine the time evolution of the disk's metallicity gradient.

We compare our old disk stars to 33 open clusters with abundance determinations from high-resolution spectra. These data are obtained from the work of four groups that have studied clusters at $R > 10$ kpc: 5 from \citet[Y05]{yon05}, 9 from Carraro et al. \citep[C07]{car04,car07,vil05,fri08}, 12 from Bragaglia, Carretta, Sestito, et al. \citep[BCS]{bra01,carretta04,carretta05,carretta07,ses06,ses07,ses08}, and 13 from Friel, Jacobson, and Pilachowski \citep[FJP]{fri05,fri10,jac08,jac09,jac11}. Seven clusters have measured abundances from more than one study, which provides an indication of the size of the uncertainties and systematics between groups. Properties of the clusters are listed in Table~\ref{clusters}. The open clusters have ages ranging from 0.6 to 10.1 Gyr; the median age is 4.0 Gyr, with only two as old as 7 Gyr, which makes this cluster sample younger than the typical age of the stars in our sample (see Table~\ref{meanages}). The ages of the clusters are indicated by the size of the symbols in Figure~\ref{r_z} and~\ref{r_feh}, with older clusters having larger symbols. For clusters with more than one measurement, a line connects the symbols representing the different determinations.

The Cepheid data are taken from \citet[AL]{and02a,and02b,and02c,and04,luc03,luc06}, and \citet[Y06]{yon06}. Most of the Cepheids in these samples are too close to the midplane ($|Z|<0.15$ kpc) to be directly compared to our sample. Those in the higher bins ($0.5 < |Z| < 1.0$ kpc), which are mostly from the \citet{yon06} sample, tend to be at larger radii than our MSTO stars. For consistency we have re-calculated $R$, $|Z|$ values using published $l$, $b$, and distances with $R_{\rm GC, \odot}=8.0$ kpc where necessary.

Both Cepheids and open clusters appear to be slightly more metal-rich than the median metallicities of the old disk stars. But systematic differences in [Fe/H] between different groups can be up to 0.2 dex, as shown by the clusters with multiple measurements (see Table~\ref{clusters} as well as the discussion in the Appendix of \citealt{fri10}). We do not see any obvious differences in metallicities between old and young clusters (large versus small open symbols). At low $|Z|$, the slopes of the open clusters and field stars do not appear to be dramatically different, as expected given the good agreement between our measurement of the radial gradient in the low $|Z|$ bins and the values of \citet{fri02} and \citet[dotted lines in Figure~\ref{slopes}]{luc06}. However, at high $|Z|$, there are three clusters with metallicities $\sim0.5$ dex higher than the median [Fe/H] for field stars at small $R$ ($< 10$ kpc). While a steeper slope in the inner regions of the disk ($-0.13$ dex kpc$^{-1}$ at $R < 8$ kpc) has been reported for both open clusters and Cepheids (e.g., \citealp{ped09,mag10}), we do not see a steeper metallicity gradient at $R < 8$ kpc in our sample of old disk stars.

The question of how the radial metallicity gradient may change over time is still debated in the literature, with conflicting reports of flattening (e.g., \citealp{fri02,che03,mac05}), steepening (e.g., \citealp{sta10}), and constant slopes (e.g., \citealp{mag09}) as a function of time. If the gradient is steeper in the inner disk, as has been reported, the flatness of the gradient for old disk stars, at face value, implies that the radial metallicity gradient in the inner disk has grown steeper with time. Radial migration complicates this issue greatly, however, as the movement of stars from their initial orbits may wash out a previously existing gradient (e.g., \citealp{ros08b}). Based on the observations presented here, we are not able to draw any clear conclusions about the time evolution of the metallicity gradient in the inner disk.

The disagreement in abundances between young and old tracers is not present in the outer disk, where the median metallicity of the old disk stars ([Fe/H] $\sim -0.5$) is consistent with the metallicities reported by \citet{yon05} for outer disk open clusters. Furthermore, the median metallicity of our sample at $|Z| > 1.0$ kpc is constant at \textit{all} values of $R$, which suggests that the flattening of the gradient is due to a trend in $|Z|$. Given the change in the radial range spanned in each $|Z|$ bin, a superposition of the negative gradient at small $|Z|$ with a flat gradient at large $|Z|$ could result in an apparent discontinuity in the radial gradient.

This result suggests that it is important to distinguish between trends in the \textit{radial} and \textit{vertical} directions. As shown in Figure~\ref{r_z}, all of the outer disk clusters from the literature ($R > 15$ kpc) are located far from the Galactic plane, and their mean metallicity is consistent with our field stars located at similar vertical heights ($|Z| > 1.0$ kpc), although our field star sample extends only to $R\sim15$ kpc, whereas the clusters reach radial distances of 20 kpc or more. Based on our observations of old disk stars, whether the reported discontinuity is truly a feature of the \textit{radial} metallicity gradient is unclear. 

Our sample does not reach to sufficiently large radii at low latitudes; with the limited range in $R$ at low $|Z|$, we cannot confirm that the trend is purely vertical. If the radial gradients we measure at $|Z| < 1.0$ kpc extend to large $R$ ($>15$kpc) \textit{for all $|Z|$}, then the discontinuity in the radial metallicity gradient of the disk actually reflects the change in slope at different $|Z|$. \citet{fri10} have stressed the need for more high-resolution data of open clusters to build up a homogeneous, statistical sample; our work shows that it is also necessary to fully sample a range of both $R$ and $|Z|$ to understand the metallicity trends in the disk. The large $R$, low $|Z|$ region of the Galaxy will be probed by future surveys such as APOGEE \citep{eis11}.

\subsection{Implications for Thick Disk Formation}
The lack of a radial metallicity gradient far from the Galactic plane provides an observational constraint that must be matched by any viable scenario for thick disk formation, such as the four described in \S\ref{intro}. A flat gradient, as we measure, may come most naturally out of a turbulent disk at high redshift (scenario 3). If the thick disk formed rapidly at early times (e.g., \citealt{bro04,bro05}), then the thick disk would be chemically homogeneous and the radial metallicity gradient far from the plane of the disk would be flat. 

The observations can only be explained by current models of radial migration in isolated disks (e.g., \citealp{ros08b,sch09a}) if radial mixing is strong (scenario 4). These models show that dynamical interactions with spiral arms can move stars from their initial orbits and make the gradient shallow, but the mechanism must be efficient if it is the dominant mechanism for forming the thick disk. The degree of radial mixing may be increased through dynamical interactions with the Galactic bar \citep{fri94,mar94,min10}.

Another possible way to increase the amount of radial mixing in a disk is to place the disk within a cosmologically motivated accretion history (scenario 1). The simulations of \citet{bir11} show that stars can move far from their initial orbits in a disk that is bombarded by multiple minor mergers (e.g., \citealt{kaz08}). Furthermore, the bombarded disk experiences more radial mixing at high $|Z|$ than an identical disk in isolation. At the midplane, however, both scenarios show about the same amount of mixing. The increased radial mixing at high $|Z|$ could be responsible for the lack of a radial gradient.

With our data alone, we cannot rule out the direct accretion of stars that originally formed in satellites that have been disrupted (scenario 2). To assess whether the thick disk exhibits the predicted clumpiness, we would need to investigate the azimuthal variation within our sample or at more detailed abundances with followup observations. But observations of the orbital properties of stars in SDSS and RAVE \citep{die10,wil11} show that there are not enough high eccentricity stars to match what is seen in the simulations of \citet{aba03}. Furthermore, recent simulations \citep{rea08,vil08} predict lower numbers of stars being directly accreted during minor mergers than the 2003 simulations.

Our conclusions are consistent with those of \citet{lee11b}, who favor a cosmological formation scenario for the thick disk; they base this interpretation on correlations between rotational velocity $V_{\phi}$ and [Fe/H] and $R$. Only recently have simulations begun to include prescriptions for the formation of stars and their impact on the expected chemistry of the disk populations (e.g., \citealt{bro05,sti10,loe11}; and references therein). The measurement of the radial metallicity gradient at different heights above the plane is an additional observational constraint that can be used to test new, improved models for disk evolution and the next generation of simulations.

\section{Summary}\label{summary}
Using a sample of 7010 MSTO stars from the SEGUE survey, we measure the radial metallicity gradient $\Delta$[Fe/H]/$\Delta R$, as a function of height above the plane $|Z|$, in the Milky Way disk. Near the midplane, where our sample is dominated by the thin disk, we see a negative radial metallicity gradient that is consistent with previously published values \citep{fri02,luc06,luc11}. At large vertical heights, where we are dominated by the thick disk, the radial metallicity gradient becomes flat, consistent with previous work \citep{all06} using a sample located at larger $|Z|$. Our sample, located at low Galactic latitude, covers a larger range in $R$ at small $|Z|$ and allows us to make a direct comparison between the thin and thick disks using the same sample.

At $|Z| > 1.0$ kpc, the median metallicity of old disk stars is consistent with the open cluster metallicities reported by \citet{yon05} and others at large $R$. In addition, our sample of disk stars shows that the flat gradient at large vertical height $|Z|$ extends to small $R$. Because the outer disk clusters are all located at large $|Z|$, the reported discontinuity in the radial gradient is consistent with a transition found using tracers at small $|Z|$ to large $|Z|$. We stress that abundances need to be examined as a function of both $R$ and $|Z|$ in order to truly understand the observed trends.

In contrast to the outer disk, open clusters and Cepheids in the inner disk at high $|Z|$ have median metallicities $\sim0.5$ dex higher than old disk stars at the same $R$ and $|Z|$; thus, far from the Galactic plane, the younger tracers do not exhibit the same flat metallicity gradient that is seen in the old disk stars. Whether this is indicative of a metallicity gradient that becomes steeper with time is unclear, as radial migration may play a role in erasing a pre-existing gradient in the old disk stars. 

A flat radial metallicity gradient in the thick disk is consistent with the predictions of a gas-rich, turbulent disk at high redshift \citep{bro05,bou09}. It may also be consistent with the scenarios of radial migration \citep{ros08b,sch09b,min10} or minor mergers \citep{kaz08,bir11}, provided that mixing in the radial direction is strong. We also cannot exclude the direct accretion of stars from satellites in minor mergers \citep{aba03}. While we are not able to conclusively rule out any of these scenarios, the change in the radial gradient as a function of height above the plane is an important observational constraint for future theoretical work.

\acknowledgments
J.Y.C. would like to H. Jacobson, E. Friel, K. Schlesinger, and J. Bird for useful conversations. C.M.R. gratefully acknowledges funding from the David and Lucile Packard Foundation, and thanks the Max-Planck-Institute f\"{u}r Astronomie (MPIA), Heidelberg for hospitality. Y.S.L. and  T.C.B. acknowledge partial funding of this work from grants PHY 02-16783 and PHY 08-22648: Physics Frontier Center/Joint Institute for Nuclear Astrophysics (JINA), awarded by the U.S. National Science Foundation.

Funding for the SDSS and SDSS-II has been provided by the Alfred P. Sloan Foundation, the Participating Institutions, the National Science Foundation, the U.S. Department of Energy, the National Aeronautics and Space Administration, the Japanese Monbukagakusho, the Max Planck Society, and the Higher Education Funding Council for England. The SDSS Web Site is http://www.sdss.org/.

The SDSS is managed by the Astrophysical Research Consortium for the Participating Institutions. The Participating Institutions are the American Museum of Natural History, Astrophysical Institute Potsdam, University of Basel, University of Cambridge, Case Western Reserve University, University of Chicago, Drexel University, Fermilab, the Institute for Advanced Study, the Japan Participation Group, Johns Hopkins University, the Joint Institute for Nuclear Astrophysics, the Kavli Institute for Particle Astrophysics and Cosmology, the Korean Scientist Group, the Chinese Academy of Sciences (LAMOST), Los Alamos National Laboratory, the Max-Planck-Institute for Astronomy (MPIA), the Max-Planck-Institute for Astrophysics (MPA), New Mexico State University, Ohio State University, University of Pittsburgh, University of Portsmouth, Princeton University, the United States Naval Observatory, and the University of Washington.

Funding for SDSS-III has been provided by the Alfred P. Sloan Foundation, the Participating Institutions, the National Science Foundation, and the U.S. Department of Energy Office of Science. The SDSS-III web site is http://www.sdss3.org/.

SDSS-III is managed by the Astrophysical Research Consortium for the Participating Institutions of the SDSS-III Collaboration including the University of Arizona, the Brazilian Participation Group, Brookhaven National Laboratory, University of Cambridge, University of Florida, the French Participation Group, the German Participation Group, the Instituto de Astrofisica de Canarias, the Michigan State/Notre Dame/JINA Participation Group, Johns Hopkins University, Lawrence Berkeley National Laboratory, Max Planck Institute for Astrophysics, New Mexico State University, New York University, Ohio State University, Pennsylvania State University, University of Portsmouth, Princeton University, the Spanish Participation Group, University of Tokyo, University of Utah, Vanderbilt University, University of Virginia, University of Washington, and Yale University.

Facilities: \facility{Sloan}.

\appendix
\section{Appendix: Calculating Weights}\label{appendix}
As discussed in \S\ref{weights}, we assign weights to each of the MSTO stars to reconstruct the properties of the underlying parent population. There are three major ways in which the spectroscopic sample is different from the full population along each line of sight: (1) The photometric objects in regions with the highest extinction were not considered for spectroscopy. (2) Not all candidates for spectroscopy are observed. (3) We observe only MSTO stars using a color cut which is biased against redder metal-rich stars. 

Each star in our sample is given three weights which correspond with the three differences listed above: (1) the \textit{area weight} $W_{\rm A}$, which depends on the coverage on the plane of the sky in each line of sight; (2) the \textit{CMD weight} $W_{\rm CMD}$, which depends on the target's location in the CMD; (3) the \textit{LF weight} $W_{\rm LF}$, which depends on the target's $T_{\rm eff}$, [Fe/H], and location in the CMD. The total weight $W$ is the product of the three weights $W_{\rm A}$, $W_{\rm CMD}$, and $W_{\rm LF}$.

\subsection{Area Weight}
The \textit{area weight} $W_{\rm A}$ corrects for the area not covered by the spectroscopic survey because the 25\% most extincted photometric objects were not considered for spectroscopy. These objects were removed using the cut described in steps 2-3 of \S\ref{targets}. The righthand panel of Figure~\ref{d_gmrext} shows an example of how the missing area is distributed in the field. Since the missing area in each line of sight is slightly different, this weight is needed to ensure that every line of sight effectively probes the same volume of the Galaxy and has equal influence on the final gradient measurement. Since the dust is mainly in the foreground and our sample is primarily distant stars, we assume that the volume of the Galaxy behind the high extinction patches is the same as the rest of the volume probed by stars along the same line of sight. 

We use the extinction map of \citetalias{sch98} to calculate the area with extinction lower than the cut. $W_{\rm A}$ is the ratio of the total area (7 square degrees) to the low extinction area. We note that the angular resolution of the \citetalias{sch98} maps is 6.1 arcminutes, so that the most extincted regions are always in irregular, contiguous patches on the sky.

\subsection{Color-Magnitude Weight}
\begin{figure}[!ht]
\epsscale{1.1}
\plotone{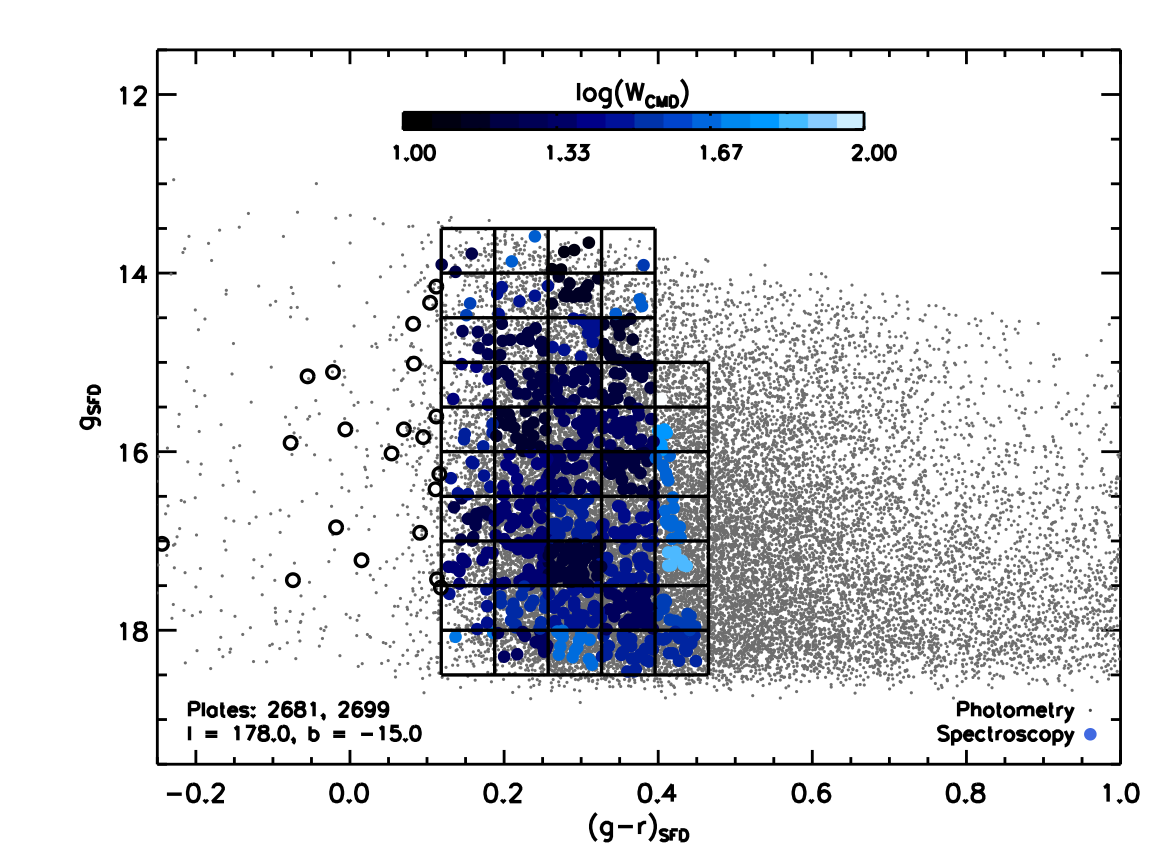}
\caption{Weighting for CMD sampling along one line of sight. For each bin (black lines) the CMD weight $W_{\rm CMD}$ is calculated by taking the ratio of the number of photometric (gray) to spectroscopic (blue) objects. Outliers bluer or redder than 2-$\sigma$ from the mean value of $(g-r)_{\rm SFD}$ are given a weight of zero (open circles). The values of $W_{\rm CMD}$ vary by less than a factor of ten, with the highest values in the reddest and faintest bins. This correction accounts for uneven sampling of the CMD and provides the normalization for the variation in the number of objects between different lines of sight.}
\label{sp_weight}
\end{figure}

The \textit{CMD weight} $W_{\rm CMD}$ normalizes between the different lines of sight; while each field has roughly the same number of spectra, the total number of photometric objects varies a great deal due to the structure of the Galaxy. This weight also accounts for any uneven sampling due to the stochastic nature of the random selection of spectroscopic targets, especially at the faint and red limits, where targets are less likely to end up in our sample because they have low quality spectra. We divide the CMD into bins of $g_{\rm SFD}$ and $(g-r)_{\rm SFD}$, as shown in Figure~\ref{sp_weight} (black lines). We use the \citetalias{sch98} colors and magnitudes because this is the CMD in which the $g-r$ cut was applied, meaning that the randomly selected spectra are an unbiased sample of the underlying CMD; this procedure does not depend on whether the \citetalias{sch98} colors and magnitudes are correct. Since the sampling is a smooth function of color and magnitude and changes slowly, the bins are sufficiently small that we can assume the sampling is constant within each bin. 

The magnitude bins are 0.5 magnitudes wide and span the entire range of the sample. Because the CMD bins use corrections from \citetalias{sch98}, which can be affected by reddening, the color bins are different for each line of sight. To determine the color bins, we calculate the mean and standard deviation of the $(g-r)_{\rm SFD}$ colors of stars in each line of sight. Spectra with colors more than 2-$\sigma$ from the mean have $W_{\rm CMD} = 0$; this removes 217 stars from the sample (open circles). Most of these are very blue objects (hotter stars that are likely not on the main sequence), although some red objects are removed as well. The remaining sample in the line of sight is divided into five equal-sized color bins. In each CMD bin, $W_{\rm CMD}$ is the ratio of the number of photometric objects (small gray dots) to the number of spectroscopic objects (blue circles) and is shown by the color coding in Figure~\ref{sp_weight}. The highest values of $W_{\rm CMD}$ are found in the reddest and faintest bins, while the variation in the middle of CMD is relatively small. The difference between the smallest and largest values is less than a factor of ten.

\subsection{Luminosity Function Weight}
The \textit{LF weight} $W_{\rm LF}$ allows us to use the MSTO sample as tracers for the total underlying population. The CMD bins used to calculate $W_{\rm LF}$ use the isochrone extinction corrected $g_0$ and $(g-r)_0$ (see \S\ref{distances}) and are different than those used to calculate $W_{\rm CMD}$. For $W_{\rm CMD}$, we had to account for the random sampling from the total photometric sample using the \citetalias{sch98} corrections, but for $W_{\rm LF}$, we use the isochrone extinction corrections because they provide the best estimates of the absolute magnitude and $g-r$ colors. The color bins are 0.1 magnitudes wide in the range $0.05 < (g-r)_0 < 0.85$, while the magnitude bins are 0.5 magnitudes wide in the range $12 < g_0 < 20$; these bins apply for all 11 lines of sight.

In each CMD bin, we find the fraction of the luminosity function that is observed in the given $(g-r)_0$ range. The weight is simply the reciprocal of the fraction. We use luminosity functions assuming a \citet{cha01} lognormal initial mass function, generated by the Padova group\footnote{\footnotesize http://stev.oapd.inaf.it/cmd} \citep{gir04}, where we have modified the faint end of the theoretical luminosity functions to more closely reflect the shapes of the luminosity functions for disk field stars reported by \citet{rei02} and \citet{boc10}. The modifications are made at magnitudes fainter than the peak of the luminosity function. The luminosity function used depends on the age and metallicity of the target. 

\begin{figure*}[!ht]
\epsscale{1.1}
\plotone{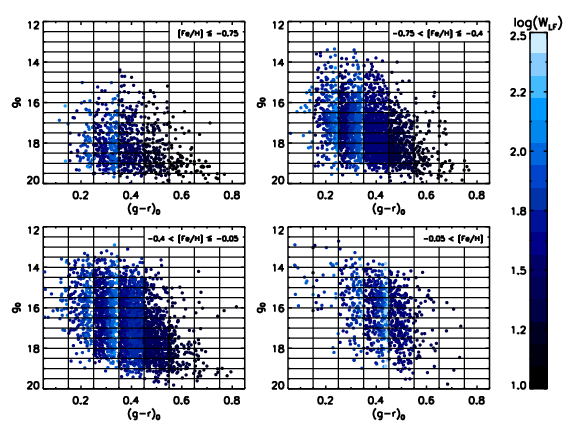}
\caption{Weighting for LF coverage for the total sample. In each isochrone extinction-corrected CMD bin (black lines) the LF weight $W_{\rm LF}$ corrects for the fraction of the luminosity function that is observed. The luminosity function used varies with both metallicity (shown in four separate panels) and age, which depends on the temperature. This correction accounts for the bias against redder, metal-rich stars that results from the $(g-r)_{\rm SFD}$ color selection.}
\label{feh_weight}
\end{figure*}

The derived values for $W_{\rm LF}$ are shown in Figure~\ref{feh_weight} for the entire sample; each panel shows a different range in metallicity. For a given $(g-r)_0$ color, $W_{\rm LF}$ is roughly constant as a function of magnitude. Within a given color bin, the change in $W_{\rm LF}$ corresponds to different main sequence turnoffs for isochrones of different ages, which are assigned based on the temperatures of the targets as described in \S\ref{distances} and Figure~\ref{isochrone_ages}. In a given CMD bin, stars with older ages have larger LF weights because a smaller fraction of the luminosity function is observable within the bin. This accounts for the vertical striping pattern seen in the four panels. We assign $W_{\rm LF} = 0$ to the three spectra that fall outside the bounds of the CMD bins.

Using the luminosity functions to correct for the fraction of unobserved stars represented by the stars in each CMD bin corrects for the bias against selecting metal-rich stars for the spectroscopic sample. The identification of MSTO stars relies on a cut in $(g-r)_{\rm SFD}$, which preferentially removes metal-rich stars that fall on isochrones with redder turnoff colors. Thus, for metal-rich stars, more of the population is removed by the color cut and a smaller fraction of the luminosity function is observed. Consequently, metal-rich targets have larger values of $W_{\rm LF}$ (lighter blue circles) at the same $(g-r)_0$ in the four panels in Figure~\ref{feh_weight}. In order for $W_{\rm LF}$ to make the proper corrections, we need to have a sufficient number of metal-rich stars in our sample to which we can apply the weights. Based on the arguments presented in \S\ref{metalbias}, our sample should satisfy this requirement.

\bibliographystyle{apj}
\bibliography{bibliography}

\begin{thebibliography}{123}
\expandafter\ifx\csname natexlab\endcsname\relax\def\natexlab#1{#1}\fi

\bibitem[{{Abadi} {et~al.}(2003){Abadi}, {Navarro}, {Steinmetz}, \&
  {Eke}}]{aba03}
{Abadi}, M.~G., {Navarro}, J.~F., {Steinmetz}, M., \& {Eke}, V.~R. 2003, \apj,
  597, 21

\bibitem[{{Abazajian} {et~al.}(2009){Abazajian}, {Adelman-McCarthy},
  {Ag{\"u}eros}, {Allam}, {Allende Prieto}, {An}, {Anderson}, {Anderson},
  {Annis}, {Bahcall}, \& et~al.}]{aba09}
{Abazajian}, K.~N., {et~al.} 2009, \apjs, 182, 543

\bibitem[{{Aihara} {et~al.}(2011){Aihara}, {Allende Prieto}, {An}, {Anderson},
  {Aubourg}, {Balbinot}, {Beers}, {Berlind}, {Bickerton}, {Bizyaev}, {Blanton},
  {Bochanski}, {Bolton}, {Bovy}, {Brandt}, {Brinkmann}, {Brown}, {Brownstein},
  {Busca}, {Campbell}, {Carr}, {Chen}, {Chiappini}, {Comparat}, {Connolly},
  {Cortes}, {Croft}, {Cuesta}, {da Costa}, {Davenport}, {Dawson}, {Dhital},
  {Ealet}, {Ebelke}, {Edmondson}, {Eisenstein}, {Escoffier}, {Esposito},
  {Evans}, {Fan}, {Femen{\'{\i}}a Castell{\'a}}, {Font-Ribera}, {Frinchaboy},
  {Ge}, {Gillespie}, {Gilmore}, {Gonz{\'a}lez Hern{\'a}ndez}, {Gott}, {Gould},
  {Grebel}, {Gunn}, {Hamilton}, {Harding}, {Harris}, {Hawley}, {Hearty}, {Ho},
  {Hogg}, {Holtzman}, {Honscheid}, {Inada}, {Ivans}, {Jiang}, {Johnson},
  {Jordan}, {Jordan}, {Kazin}, {Kirkby}, {Klaene}, {Knapp}, {Kneib},
  {Kochanek}, {Koesterke}, {Kollmeier}, {Kron}, {Lampeitl}, {Lang}, {Le Goff},
  {Lee}, {Lin}, {Long}, {Loomis}, {Lucatello}, {Lundgren}, {Lupton}, {Ma},
  {MacDonald}, {Mahadevan}, {Maia}, {Makler}, {Malanushenko}, {Malanushenko},
  {Mandelbaum}, {Maraston}, {Margala}, {Masters}, {McBride}, {McGehee},
  {McGreer}, {M{\'e}nard}, {Miralda-Escud{\'e}}, {Morrison}, {Mullally},
  {Muna}, {Munn}, {Murayama}, {Myers}, {Naugle}, {Fausti Neto}, {Cuong Nguyen},
  {Nichol}, {O'Connell}, {Ogando}, {Olmstead}, {Oravetz}, {Padmanabhan},
  {Palanque-Delabrouille}, {Pan}, {Pandey}, {P{\^a}ris}, {Percival},
  {Petitjean}, {Pfaffenberger}, {Pforr}, {Phleps}, {Pichon}, {Pieri}, {Prada},
  {Price-Whelan}, {Raddick}, {Ramos}, {Reyl{\'e}}, {Rich}, {Richards}, {Rix},
  {Robin}, {Rocha-Pinto}, {Rockosi}, {Roe}, {Rollinde}, {Ross}, {Ross},
  {Rossetto}, {S{\'a}nchez}, {Sayres}, {Schlegel}, {Schlesinger}, {Schmidt},
  {Schneider}, {Sheldon}, {Shu}, {Simmerer}, {Simmons}, {Sivarani}, {Snedden},
  {Sobeck}, {Steinmetz}, {Strauss}, {Szalay}, {Tanaka}, {Thakar}, {Thomas},
  {Tinker}, {Tofflemire}, {Tojeiro}, {Tremonti}, {Vandenberg}, {Vargas
  Maga{\~n}a}, {Verde}, {Vogt}, {Wake}, {Wang}, {Weaver}, {Weinberg}, {White},
  {White}, {Yanny}, {Yasuda}, {Yeche}, \& {Zehavi}}]{aih11}
{Aihara}, H., {et~al.} 2011, \apjs, 193, 29

\bibitem[{{Allende Prieto} {et~al.}(2006){Allende Prieto}, {Beers}, {Wilhelm},
  {Newberg}, {Rockosi}, {Yanny}, \& {Lee}}]{all06}
{Allende Prieto}, C., {Beers}, T.~C., {Wilhelm}, R., {Newberg}, H.~J.,
  {Rockosi}, C.~M., {Yanny}, B., \& {Lee}, Y.~S. 2006, \apj, 636, 804

\bibitem[{{An} {et~al.}(2009){An}, {Pinsonneault}, {Masseron}, {Delahaye},
  {Johnson}, {Terndrup}, {Beers}, {Ivans}, \& {Ivezi{\'c}}}]{an09}
{An}, D., {et~al.} 2009, \apj, 700, 523

\bibitem[{{Andrievsky} {et~al.}(2002{\natexlab{a}}){Andrievsky}, {Bersier},
  {Kovtyukh}, {Luck}, {Maciel}, {L{\'e}pine}, \& {Beletsky}}]{and02b}
{Andrievsky}, S.~M., {Bersier}, D., {Kovtyukh}, V.~V., {Luck}, R.~E., {Maciel},
  W.~J., {L{\'e}pine}, J.~R.~D., \& {Beletsky}, Y.~V. 2002{\natexlab{a}}, \aap,
  384, 140

\bibitem[{{Andrievsky} {et~al.}(2002{\natexlab{b}}){Andrievsky}, {Kovtyukh},
  {Luck}, {L{\'e}pine}, {Maciel}, \& {Beletsky}}]{and02c}
{Andrievsky}, S.~M., {Kovtyukh}, V.~V., {Luck}, R.~E., {L{\'e}pine}, J.~R.~D.,
  {Maciel}, W.~J., \& {Beletsky}, Y.~V. 2002{\natexlab{b}}, \aap, 392, 491

\bibitem[{{Andrievsky} {et~al.}(2004){Andrievsky}, {Luck}, {Martin}, \&
  {L{\'e}pine}}]{and04}
{Andrievsky}, S.~M., {Luck}, R.~E., {Martin}, P., \& {L{\'e}pine}, J.~R.~D.
  2004, \aap, 413, 159

\bibitem[{{Andrievsky} {et~al.}(2002{\natexlab{c}}){Andrievsky}, {Kovtyukh},
  {Luck}, {L{\'e}pine}, {Bersier}, {Maciel}, {Barbuy}, {Klochkova}, {Panchuk},
  \& {Karpischek}}]{and02a}
{Andrievsky}, S.~M., {et~al.} 2002{\natexlab{c}}, \aap, 381, 32

\bibitem[{{Belokurov} {et~al.}(2007){Belokurov}, {Evans}, {Irwin},
  {Lynden-Bell}, {Yanny}, {Vidrih}, {Gilmore}, {Seabroke}, {Zucker},
  {Wilkinson}, {Hewett}, {Bramich}, {Fellhauer}, {Newberg}, {Wyse}, {Beers},
  {Bell}, {Barentine}, {Brinkmann}, {Cole}, {Pan}, \& {York}}]{bel07}
{Belokurov}, V., {et~al.} 2007, \apj, 658, 337

\bibitem[{{Bensby} {et~al.}(2003){Bensby}, {Feltzing}, \&
  {Lundstr{\"o}m}}]{ben03}
{Bensby}, T., {Feltzing}, S., \& {Lundstr{\"o}m}, I. 2003, \aap, 410, 527

\bibitem[{{Bensby} {et~al.}(2004){Bensby}, {Feltzing}, \&
  {Lundstr{\"o}m}}]{ben04b}
---. 2004, \aap, 421, 969

\bibitem[{{Bensby} {et~al.}(2005){Bensby}, {Feltzing}, {Lundstr{\"o}m}, \&
  {Ilyin}}]{ben05}
{Bensby}, T., {Feltzing}, S., {Lundstr{\"o}m}, I., \& {Ilyin}, I. 2005, \aap,
  433, 185

\bibitem[{{Bird} {et~al.}(2011){Bird}, {Kazantzidis}, \& {Weinberg}}]{bir11}
{Bird}, J.~C., {Kazantzidis}, S., \& {Weinberg}, D.~H. 2011, ArXiv e-prints

\bibitem[{{Bochanski} {et~al.}(2010){Bochanski}, {Hawley}, {Covey}, {West},
  {Reid}, {Golimowski}, \& {Ivezi{\'c}}}]{boc10}
{Bochanski}, J.~J., {Hawley}, S.~L., {Covey}, K.~R., {West}, A.~A., {Reid},
  I.~N., {Golimowski}, D.~A., \& {Ivezi{\'c}}, {\v Z}. 2010, \aj, 139, 2679

\bibitem[{{Bournaud} {et~al.}(2009){Bournaud}, {Elmegreen}, \&
  {Martig}}]{bou09}
{Bournaud}, F., {Elmegreen}, B.~G., \& {Martig}, M. 2009, \apjl, 707, L1

\bibitem[{{Bragaglia} {et~al.}(2008){Bragaglia}, {Sestito}, {Villanova},
  {Carretta}, {Randich}, \& {Tosi}}]{bra08}
{Bragaglia}, A., {Sestito}, P., {Villanova}, S., {Carretta}, E., {Randich}, S.,
  \& {Tosi}, M. 2008, \aap, 480, 79

\bibitem[{{Bragaglia} {et~al.}(2001){Bragaglia}, {Carretta}, {Gratton}, {Tosi},
  {Bonanno}, {Bruno}, {Cal{\`\i}}, {Claudi}, {Cosentino}, {Desidera},
  {Farisato}, {Rebeschini}, \& {Scuderi}}]{bra01}
{Bragaglia}, A., {et~al.} 2001, \aj, 121, 327

\bibitem[{{Brook} {et~al.}(2005){Brook}, {Gibson}, {Martel}, \&
  {Kawata}}]{bro05}
{Brook}, C.~B., {Gibson}, B.~K., {Martel}, H., \& {Kawata}, D. 2005, \apj, 630,
  298

\bibitem[{{Brook} {et~al.}(2004){Brook}, {Kawata}, {Gibson}, \&
  {Freeman}}]{bro04}
{Brook}, C.~B., {Kawata}, D., {Gibson}, B.~K., \& {Freeman}, K.~C. 2004, \apj,
  612, 894

\bibitem[{{Brooks} {et~al.}(2009){Brooks}, {Governato}, {Quinn}, {Brook}, \&
  {Wadsley}}]{bro09}
{Brooks}, A.~M., {Governato}, F., {Quinn}, T., {Brook}, C.~B., \& {Wadsley}, J.
  2009, \apj, 694, 396

\bibitem[{{Caputo} {et~al.}(2001){Caputo}, {Marconi}, {Musella}, \&
  {Pont}}]{cap01}
{Caputo}, F., {Marconi}, M., {Musella}, I., \& {Pont}, F. 2001, \aap, 372, 544

\bibitem[{{Carraro} {et~al.}(2004){Carraro}, {Bresolin}, {Villanova},
  {Matteucci}, {Patat}, \& {Romaniello}}]{car04}
{Carraro}, G., {Bresolin}, F., {Villanova}, S., {Matteucci}, F., {Patat}, F.,
  \& {Romaniello}, M. 2004, \aj, 128, 1676

\bibitem[{{Carraro} {et~al.}(2007){Carraro}, {Geisler}, {Villanova},
  {Frinchaboy}, \& {Majewski}}]{car07}
{Carraro}, G., {Geisler}, D., {Villanova}, S., {Frinchaboy}, P.~M., \&
  {Majewski}, S.~R. 2007, \aap, 476, 217

\bibitem[{{Carretta} {et~al.}(2007){Carretta}, {Bragaglia}, \&
  {Gratton}}]{carretta07}
{Carretta}, E., {Bragaglia}, A., \& {Gratton}, R.~G. 2007, \aap, 473, 129

\bibitem[{{Carretta} {et~al.}(2004){Carretta}, {Bragaglia}, {Gratton}, \&
  {Tosi}}]{carretta04}
{Carretta}, E., {Bragaglia}, A., {Gratton}, R.~G., \& {Tosi}, M. 2004, \aap,
  422, 951

\bibitem[{{Carretta} {et~al.}(2005){Carretta}, {Bragaglia}, {Gratton}, \&
  {Tosi}}]{carretta05}
---. 2005, \aap, 441, 131

\bibitem[{{Casagrande} {et~al.}(2011){Casagrande}, {Sch{\"o}nrich}, {Asplund},
  {Cassisi}, {Ram{\'{\i}}rez}, {Mel{\'e}ndez}, {Bensby}, \& {Feltzing}}]{cas11}
{Casagrande}, L., {Sch{\"o}nrich}, R., {Asplund}, M., {Cassisi}, S.,
  {Ram{\'{\i}}rez}, I., {Mel{\'e}ndez}, J., {Bensby}, T., \& {Feltzing}, S.
  2011, \aap, 530, A138+

\bibitem[{{Cescutti} {et~al.}(2007){Cescutti}, {Matteucci}, {Fran{\c c}ois}, \&
  {Chiappini}}]{ces07}
{Cescutti}, G., {Matteucci}, F., {Fran{\c c}ois}, P., \& {Chiappini}, C. 2007,
  \aap, 462, 943

\bibitem[{{Chabrier}(2001)}]{cha01}
{Chabrier}, G. 2001, \apj, 554, 1274

\bibitem[{{Chen} {et~al.}(2003){Chen}, {Hou}, \& {Wang}}]{che03}
{Chen}, L., {Hou}, J.~L., \& {Wang}, J.~J. 2003, \aj, 125, 1397

\bibitem[{{Chiappini} {et~al.}(1997){Chiappini}, {Matteucci}, \&
  {Gratton}}]{chi97}
{Chiappini}, C., {Matteucci}, F., \& {Gratton}, R. 1997, \apj, 477, 765

\bibitem[{{Chiappini} {et~al.}(2001){Chiappini}, {Matteucci}, \&
  {Romano}}]{chi01}
{Chiappini}, C., {Matteucci}, F., \& {Romano}, D. 2001, \apj, 554, 1044

\bibitem[{{Chiba} \& {Beers}(2000)}]{chi00}
{Chiba}, M., \& {Beers}, T.~C. 2000, \aj, 119, 2843

\bibitem[{{Cresci} {et~al.}(2010){Cresci}, {Mannucci}, {Maiolino}, {Marconi},
  {Gnerucci}, \& {Magrini}}]{cre10}
{Cresci}, G., {Mannucci}, F., {Maiolino}, R., {Marconi}, A., {Gnerucci}, A., \&
  {Magrini}, L. 2010, \nat, 467, 811

\bibitem[{{Dalcanton} \& {Bernstein}(2002)}]{dal02}
{Dalcanton}, J.~J., \& {Bernstein}, R.~A. 2002, \aj, 124, 1328

\bibitem[{{de Jong} {et~al.}(2010){de Jong}, {Yanny}, {Rix}, {Dolphin},
  {Martin}, \& {Beers}}]{dej10}
{de Jong}, J.~T.~A., {Yanny}, B., {Rix}, H., {Dolphin}, A.~E., {Martin}, N.~F.,
  \& {Beers}, T.~C. 2010, \apj, 714, 663

\bibitem[{{Dekel} {et~al.}(2009){Dekel}, {Sari}, \& {Ceverino}}]{dek09}
{Dekel}, A., {Sari}, R., \& {Ceverino}, D. 2009, \apj, 703, 785

\bibitem[{{Di Matteo} {et~al.}(2011){Di Matteo}, {Lehnert}, {Qu}, \& {van
  Driel}}]{di-11}
{Di Matteo}, P., {Lehnert}, M.~D., {Qu}, Y., \& {van Driel}, W. 2011, \aap,
  525, L3+

\bibitem[{{Dierickx} {et~al.}(2010){Dierickx}, {Klement}, {Rix}, \&
  {Liu}}]{die10}
{Dierickx}, M., {Klement}, R., {Rix}, H.-W., \& {Liu}, C. 2010, \apjl, 725,
  L186

\bibitem[{{Dotter} {et~al.}(2008){Dotter}, {Chaboyer}, {Jevremovi{\'c}},
  {Kostov}, {Baron}, \& {Ferguson}}]{dot08}
{Dotter}, A., {Chaboyer}, B., {Jevremovi{\'c}}, D., {Kostov}, V., {Baron}, E.,
  \& {Ferguson}, J.~W. 2008, \apjs, 178, 89

\bibitem[{{Eisenstein} {et~al.}(2011){Eisenstein}, {Weinberg}, {Agol},
  {Aihara}, {Allende Prieto}, {Anderson}, {Arns}, {Aubourg}, {Bailey},
  {Balbinot}, \& et~al.}]{eis11}
{Eisenstein}, D.~J., {et~al.} 2011, \aj, 142, 72

\bibitem[{{Elmegreen} \& {Elmegreen}(2005)}]{elm05}
{Elmegreen}, B.~G., \& {Elmegreen}, D.~M. 2005, \apj, 627, 632

\bibitem[{{Elmegreen} \& {Elmegreen}(2006)}]{elm06}
---. 2006, \apj, 650, 644

\bibitem[{{Friedli} {et~al.}(1994){Friedli}, {Benz}, \& {Kennicutt}}]{fri94}
{Friedli}, D., {Benz}, W., \& {Kennicutt}, R. 1994, \apjl, 430, L105

\bibitem[{{Friel} {et~al.}(2005){Friel}, {Jacobson}, \& {Pilachowski}}]{fri05}
{Friel}, E.~D., {Jacobson}, H.~R., \& {Pilachowski}, C.~A. 2005, \aj, 129, 2725

\bibitem[{{Friel} {et~al.}(2010){Friel}, {Jacobson}, \& {Pilachowski}}]{fri10}
---. 2010, \aj, 139, 1942

\bibitem[{{Friel} {et~al.}(2002){Friel}, {Janes}, {Tavarez}, {Scott},
  {Katsanis}, {Lotz}, {Hong}, \& {Miller}}]{fri02}
{Friel}, E.~D., {Janes}, K.~A., {Tavarez}, M., {Scott}, J., {Katsanis}, R.,
  {Lotz}, J., {Hong}, L., \& {Miller}, N. 2002, \aj, 124, 2693

\bibitem[{{Frinchaboy} {et~al.}(2008){Frinchaboy}, {Marino}, {Villanova},
  {Carraro}, {Majewski}, \& {Geisler}}]{fri08}
{Frinchaboy}, P.~M., {Marino}, A.~F., {Villanova}, S., {Carraro}, G.,
  {Majewski}, S.~R., \& {Geisler}, D. 2008, \mnras, 391, 39

\bibitem[{{Fukugita} {et~al.}(1996){Fukugita}, {Ichikawa}, {Gunn}, {Doi},
  {Shimasaku}, \& {Schneider}}]{fuk96}
{Fukugita}, M., {Ichikawa}, T., {Gunn}, J.~E., {Doi}, M., {Shimasaku}, K., \&
  {Schneider}, D.~P. 1996, \aj, 111, 1748

\bibitem[{{Gilmore} \& {Reid}(1983)}]{gil83}
{Gilmore}, G., \& {Reid}, N. 1983, \mnras, 202, 1025

\bibitem[{{Gilmore} {et~al.}(1995){Gilmore}, {Wyse}, \& {Jones}}]{gil95}
{Gilmore}, G., {Wyse}, R.~F.~G., \& {Jones}, J.~B. 1995, \aj, 109, 1095

\bibitem[{{Girardi} {et~al.}(2004){Girardi}, {Grebel}, {Odenkirchen}, \&
  {Chiosi}}]{gir04}
{Girardi}, L., {Grebel}, E.~K., {Odenkirchen}, M., \& {Chiosi}, C. 2004, \aap,
  422, 205

\bibitem[{{Goetz} \& {Koeppen}(1992)}]{goe92}
{Goetz}, M., \& {Koeppen}, J. 1992, \aap, 262, 455

\bibitem[{{Gunn} {et~al.}(1998){Gunn}, {Carr}, {Rockosi}, {Sekiguchi}, {Berry},
  {Elms}, {de Haas}, {Ivezi{\'c}}, {Knapp}, {Lupton}, {Pauls}, {Simcoe},
  {Hirsch}, {Sanford}, {Wang}, {York}, {Harris}, {Annis}, {Bartozek},
  {Boroski}, {Bakken}, {Haldeman}, {Kent}, {Holm}, {Holmgren}, {Petravick},
  {Prosapio}, {Rechenmacher}, {Doi}, {Fukugita}, {Shimasaku}, {Okada}, {Hull},
  {Siegmund}, {Mannery}, {Blouke}, {Heidtman}, {Schneider}, {Lucinio}, \&
  {Brinkman}}]{gun98}
{Gunn}, J.~E., {et~al.} 1998, \aj, 116, 3040

\bibitem[{{Gunn} {et~al.}(2006){Gunn}, {Siegmund}, {Mannery}, {Owen}, {Hull},
  {Leger}, {Carey}, {Knapp}, {York}, {Boroski}, {Kent}, {Lupton}, {Rockosi},
  {Evans}, {Waddell}, {Anderson}, {Annis}, {Barentine}, {Bartoszek}, {Bastian},
  {Bracker}, {Brewington}, {Briegel}, {Brinkmann}, {Brown}, {Carr},
  {Czarapata}, {Drennan}, {Dombeck}, {Federwitz}, {Gillespie}, {Gonzales},
  {Hansen}, {Harvanek}, {Hayes}, {Jordan}, {Kinney}, {Klaene}, {Kleinman},
  {Kron}, {Kresinski}, {Lee}, {Limmongkol}, {Lindenmeyer}, {Long}, {Loomis},
  {McGehee}, {Mantsch}, {Neilsen}, {Neswold}, {Newman}, {Nitta}, {Peoples},
  {Pier}, {Prieto}, {Prosapio}, {Rivetta}, {Schneider}, {Snedden}, \&
  {Wang}}]{gun06}
---. 2006, \aj, 131, 2332

\bibitem[{{Hayashi} \& {Chiba}(2006)}]{hay06}
{Hayashi}, H., \& {Chiba}, M. 2006, \pasj, 58, 835

\bibitem[{{Haywood}(2008)}]{hay08}
{Haywood}, M. 2008, \mnras, 388, 1175

\bibitem[{{Ibata} {et~al.}(2001){Ibata}, {Irwin}, {Lewis}, {Ferguson}, \&
  {Tanvir}}]{iba01}
{Ibata}, R., {Irwin}, M., {Lewis}, G., {Ferguson}, A.~M.~N., \& {Tanvir}, N.
  2001, \nat, 412, 49

\bibitem[{{Ivezi{\'c}} {et~al.}(2008){Ivezi{\'c}}, {Sesar}, {Juri{\'c}},
  {Bond}, {Dalcanton}, {Rockosi}, {Yanny}, {Newberg}, {Beers}, {Allende
  Prieto}, {Wilhelm}, {Lee}, {Sivarani}, {Norris}, {Bailer-Jones}, {Re
  Fiorentin}, {Schlegel}, {Uomoto}, {Lupton}, {Knapp}, {Gunn}, {Covey},
  {Smith}, {Miknaitis}, {Doi}, {Tanaka}, {Fukugita}, {Kent}, {Finkbeiner},
  {Munn}, {Pier}, {Quinn}, {Hawley}, {Anderson}, {Kiuchi}, {Chen}, {Bushong},
  {Sohi}, {Haggard}, {Kimball}, {Barentine}, {Brewington}, {Harvanek},
  {Kleinman}, {Krzesinski}, {Long}, {Nitta}, {Snedden}, {Lee}, {Harris},
  {Brinkmann}, {Schneider}, \& {York}}]{ive08}
{Ivezi{\'c}}, {\v Z}., {et~al.} 2008, \apj, 684, 287

\bibitem[{{Jacobson} {et~al.}(2008){Jacobson}, {Friel}, \&
  {Pilachowski}}]{jac08}
{Jacobson}, H.~R., {Friel}, E.~D., \& {Pilachowski}, C.~A. 2008, \aj, 135, 2341

\bibitem[{{Jacobson} {et~al.}(2009){Jacobson}, {Friel}, \&
  {Pilachowski}}]{jac09}
---. 2009, \aj, 137, 4753

\bibitem[{{Jacobson} {et~al.}(2011){Jacobson}, {Friel}, \&
  {Pilachowski}}]{jac11}
---. 2011, \aj, 141, 58

\bibitem[{{Juri{\'c}} {et~al.}(2008){Juri{\'c}}, {Ivezi{\'c}}, {Brooks},
  {Lupton}, {Schlegel}, {Finkbeiner}, {Padmanabhan}, {Bond}, {Sesar},
  {Rockosi}, {Knapp}, {Gunn}, {Sumi}, {Schneider}, {Barentine}, {Brewington},
  {Brinkmann}, {Fukugita}, {Harvanek}, {Kleinman}, {Krzesinski}, {Long},
  {Neilsen}, {Nitta}, {Snedden}, \& {York}}]{jur08}
{Juri{\'c}}, M., {et~al.} 2008, \apj, 673, 864

\bibitem[{{Kazantzidis} {et~al.}(2008){Kazantzidis}, {Bullock}, {Zentner},
  {Kravtsov}, \& {Moustakas}}]{kaz08}
{Kazantzidis}, S., {Bullock}, J.~S., {Zentner}, A.~R., {Kravtsov}, A.~V., \&
  {Moustakas}, L.~A. 2008, \apj, 688, 254

\bibitem[{{Kazantzidis} {et~al.}(2009){Kazantzidis}, {Zentner}, {Kravtsov},
  {Bullock}, \& {Debattista}}]{kaz09}
{Kazantzidis}, S., {Zentner}, A.~R., {Kravtsov}, A.~V., {Bullock}, J.~S., \&
  {Debattista}, V.~P. 2009, \apj, 700, 1896

\bibitem[{{Lacey} \& {Fall}(1985)}]{lac85}
{Lacey}, C.~G., \& {Fall}, S.~M. 1985, \apj, 290, 154

\bibitem[{{Larson}(1976)}]{lar76}
{Larson}, R.~B. 1976, \mnras, 176, 31

\bibitem[{{Lee} {et~al.}(2008{\natexlab{a}}){Lee}, {Beers}, {Sivarani},
  {Allende Prieto}, {Koesterke}, {Wilhelm}, {Re Fiorentin}, {Bailer-Jones},
  {Norris}, {Rockosi}, {Yanny}, {Newberg}, {Covey}, {Zhang}, \& {Luo}}]{lee08a}
{Lee}, Y.~S., {et~al.} 2008{\natexlab{a}}, \aj, 136, 2022

\bibitem[{{Lee} {et~al.}(2008{\natexlab{b}}){Lee}, {Beers}, {Sivarani},
  {Johnson}, {An}, {Wilhelm}, {Allende Prieto}, {Koesterke}, {Re Fiorentin},
  {Bailer-Jones}, {Norris}, {Yanny}, {Rockosi}, {Newberg}, {Cudworth}, \&
  {Pan}}]{lee08b}
---. 2008{\natexlab{b}}, \aj, 136, 2050

\bibitem[{{Lee} {et~al.}(2011{\natexlab{a}}){Lee}, {Beers}, {An}, {Ivezi{\'c}},
  {Just}, {Rockosi}, {Morrison}, {Johnson}, {Sch{\"o}nrich}, {Bird}, {Yanny},
  {Harding}, \& {Rocha-Pinto}}]{lee11b}
---. 2011{\natexlab{a}}, \apj, 738, 187

\bibitem[{{Lee} {et~al.}(2011{\natexlab{b}}){Lee}, {Beers}, {Allende Prieto},
  {Lai}, {Rockosi}, {Morrison}, {Johnson}, {An}, {Sivarani}, \&
  {Yanny}}]{lee11a}
---. 2011{\natexlab{b}}, \aj, 141, 90

\bibitem[{{Lemasle} {et~al.}(2008){Lemasle}, {Fran{\c c}ois}, {Piersimoni},
  {Pedicelli}, {Bono}, {Laney}, {Primas}, \& {Romaniello}}]{lem08}
{Lemasle}, B., {Fran{\c c}ois}, P., {Piersimoni}, A., {Pedicelli}, S., {Bono},
  G., {Laney}, C.~D., {Primas}, F., \& {Romaniello}, M. 2008, \aap, 490, 613

\bibitem[{{Loebman} {et~al.}(2011){Loebman}, {Ro{\v s}kar}, {Debattista},
  {Ivezi{\'c}}, {Quinn}, \& {Wadsley}}]{loe11}
{Loebman}, S.~R., {Ro{\v s}kar}, R., {Debattista}, V.~P., {Ivezi{\'c}}, {\v
  Z}., {Quinn}, T.~R., \& {Wadsley}, J. 2011, \apj, 737, 8

\bibitem[{{Luck} {et~al.}(2003){Luck}, {Gieren}, {Andrievsky}, {Kovtyukh},
  {Fouqu{\'e}}, {Pont}, \& {Kienzle}}]{luc03}
{Luck}, R.~E., {Gieren}, W.~P., {Andrievsky}, S.~M., {Kovtyukh}, V.~V.,
  {Fouqu{\'e}}, P., {Pont}, F., \& {Kienzle}, F. 2003, \aap, 401, 939

\bibitem[{{Luck} {et~al.}(2006){Luck}, {Kovtyukh}, \& {Andrievsky}}]{luc06}
{Luck}, R.~E., {Kovtyukh}, V.~V., \& {Andrievsky}, S.~M. 2006, \aj, 132, 902

\bibitem[{{Luck} \& {Lambert}(2011)}]{luc11}
{Luck}, R.~E., \& {Lambert}, D.~L. 2011, \aj, 142, 136

\bibitem[{{Maciel} {et~al.}(2005){Maciel}, {Lago}, \& {Costa}}]{mac05}
{Maciel}, W.~J., {Lago}, L.~G., \& {Costa}, R.~D.~D. 2005, \aap, 433, 127

\bibitem[{{Magrini} {et~al.}(2010){Magrini}, {Randich}, {Zoccali}, {Jilkova},
  {Carraro}, {Galli}, {Maiorca}, \& {Busso}}]{mag10}
{Magrini}, L., {Randich}, S., {Zoccali}, M., {Jilkova}, L., {Carraro}, G.,
  {Galli}, D., {Maiorca}, E., \& {Busso}, M. 2010, \aap, 523, A11+

\bibitem[{{Magrini} {et~al.}(2009){Magrini}, {Sestito}, {Randich}, \&
  {Galli}}]{mag09}
{Magrini}, L., {Sestito}, P., {Randich}, S., \& {Galli}, D. 2009, \aap, 494, 95

\bibitem[{{Martin} \& {Roy}(1994)}]{mar94}
{Martin}, P., \& {Roy}, J.-R. 1994, \apj, 424, 599

\bibitem[{{Mart{\'{\i}}nez-Delgado} {et~al.}(2010){Mart{\'{\i}}nez-Delgado},
  {Gabany}, {Crawford}, {Zibetti}, {Majewski}, {Rix}, {Fliri},
  {Carballo-Bello}, {Bardalez-Gagliuffi}, {Pe{\~n}arrubia}, {Chonis}, {Madore},
  {Trujillo}, {Schirmer}, \& {McDavid}}]{mar10}
{Mart{\'{\i}}nez-Delgado}, D., {et~al.} 2010, \aj, 140, 962

\bibitem[{{Matteucci} \& {Francois}(1989)}]{mat89}
{Matteucci}, F., \& {Francois}, P. 1989, \mnras, 239, 885

\bibitem[{{Minchev} \& {Famaey}(2010)}]{min10}
{Minchev}, I., \& {Famaey}, B. 2010, \apj, 722, 112

\bibitem[{{Newberg} {et~al.}(2002){Newberg}, {Yanny}, {Rockosi}, {Grebel},
  {Rix}, {Brinkmann}, {Csabai}, {Hennessy}, {Hindsley}, {Ibata}, {Ivezi{\'c}},
  {Lamb}, {Nash}, {Odenkirchen}, {Rave}, {Schneider}, {Smith}, {Stolte}, \&
  {York}}]{new02}
{Newberg}, H.~J., {et~al.} 2002, \apj, 569, 245

\bibitem[{{Nordstr{\"o}m} {et~al.}(2004){Nordstr{\"o}m}, {Mayor}, {Andersen},
  {Holmberg}, {Pont}, {J{\o}rgensen}, {Olsen}, {Udry}, \& {Mowlavi}}]{nor04}
{Nordstr{\"o}m}, B., {et~al.} 2004, \aap, 418, 989

\bibitem[{{Pedicelli} {et~al.}(2009){Pedicelli}, {Bono}, {Lemasle}, {Fran{\c
  c}ois}, {Groenewegen}, {Lub}, {Pel}, {Laney}, {Piersimoni}, {Romaniello},
  {Buonanno}, {Caputo}, {Cassisi}, {Castelli}, {Leurini}, {Pietrinferni},
  {Primas}, \& {Pritchard}}]{ped09}
{Pedicelli}, S., {et~al.} 2009, \aap, 504, 81

\bibitem[{{Portinari} \& {Chiosi}(2000)}]{por00}
{Portinari}, L., \& {Chiosi}, C. 2000, \aap, 355, 929

\bibitem[{{Prantzos} \& {Boissier}(2000)}]{pra00}
{Prantzos}, N., \& {Boissier}, S. 2000, \mnras, 313, 338

\bibitem[{{Purcell} {et~al.}(2009){Purcell}, {Kazantzidis}, \&
  {Bullock}}]{pur09}
{Purcell}, C.~W., {Kazantzidis}, S., \& {Bullock}, J.~S. 2009, \apjl, 694, L98

\bibitem[{{Randich} {et~al.}(2006){Randich}, {Sestito}, {Primas},
  {Pallavicini}, \& {Pasquini}}]{ran06}
{Randich}, S., {Sestito}, P., {Primas}, F., {Pallavicini}, R., \& {Pasquini},
  L. 2006, \aap, 450, 557

\bibitem[{{Read} {et~al.}(2008){Read}, {Lake}, {Agertz}, \&
  {Debattista}}]{rea08}
{Read}, J.~I., {Lake}, G., {Agertz}, O., \& {Debattista}, V.~P. 2008, \mnras,
  389, 1041

\bibitem[{{Reid} {et~al.}(2002){Reid}, {Gizis}, \& {Hawley}}]{rei02}
{Reid}, I.~N., {Gizis}, J.~E., \& {Hawley}, S.~L. 2002, \aj, 124, 2721

\bibitem[{{Ro{\v s}kar} {et~al.}(2008{\natexlab{a}}){Ro{\v s}kar},
  {Debattista}, {Quinn}, {Stinson}, \& {Wadsley}}]{ros08b}
{Ro{\v s}kar}, R., {Debattista}, V.~P., {Quinn}, T.~R., {Stinson}, G.~S., \&
  {Wadsley}, J. 2008{\natexlab{a}}, \apjl, 684, L79

\bibitem[{{Ro{\v s}kar} {et~al.}(2008{\natexlab{b}}){Ro{\v s}kar},
  {Debattista}, {Stinson}, {Quinn}, {Kaufmann}, \& {Wadsley}}]{ros08a}
{Ro{\v s}kar}, R., {Debattista}, V.~P., {Stinson}, G.~S., {Quinn}, T.~R.,
  {Kaufmann}, T., \& {Wadsley}, J. 2008{\natexlab{b}}, \apjl, 675, L65

\bibitem[{{Sales} {et~al.}(2009){Sales}, {Helmi}, {Abadi}, {Brook},
  {G{\'o}mez}, {Ro{\v s}kar}, {Debattista}, {House}, {Steinmetz}, \&
  {Villalobos}}]{sal09a}
{Sales}, L.~V., {et~al.} 2009, \mnras, 400, L61

\bibitem[{{S{\'a}nchez-Bl{\'a}zquez} {et~al.}(2009){S{\'a}nchez-Bl{\'a}zquez},
  {Courty}, {Gibson}, \& {Brook}}]{san09}
{S{\'a}nchez-Bl{\'a}zquez}, P., {Courty}, S., {Gibson}, B.~K., \& {Brook},
  C.~B. 2009, \mnras, 398, 591

\bibitem[{{Schlegel} {et~al.}(1998){Schlegel}, {Finkbeiner}, \&
  {Davis}}]{sch98}
{Schlegel}, D.~J., {Finkbeiner}, D.~P., \& {Davis}, M. 1998, \apj, 500, 525

\bibitem[{{Sch{\"o}nrich} \& {Binney}(2009{\natexlab{a}})}]{sch09a}
{Sch{\"o}nrich}, R., \& {Binney}, J. 2009{\natexlab{a}}, \mnras, 396, 203

\bibitem[{{Sch{\"o}nrich} \& {Binney}(2009{\natexlab{b}})}]{sch09b}
---. 2009{\natexlab{b}}, \mnras, 399, 1145

\bibitem[{{Sellwood} \& {Binney}(2002)}]{sel02}
{Sellwood}, J.~A., \& {Binney}, J.~J. 2002, \mnras, 336, 785

\bibitem[{{Sestito} {et~al.}(2006){Sestito}, {Bragaglia}, {Randich},
  {Carretta}, {Prisinzano}, \& {Tosi}}]{ses06}
{Sestito}, P., {Bragaglia}, A., {Randich}, S., {Carretta}, E., {Prisinzano},
  L., \& {Tosi}, M. 2006, \aap, 458, 121

\bibitem[{{Sestito} {et~al.}(2008){Sestito}, {Bragaglia}, {Randich},
  {Pallavicini}, {Andrievsky}, \& {Korotin}}]{ses08}
{Sestito}, P., {Bragaglia}, A., {Randich}, S., {Pallavicini}, R., {Andrievsky},
  S.~M., \& {Korotin}, S.~A. 2008, \aap, 488, 943

\bibitem[{{Sestito} {et~al.}(2007){Sestito}, {Randich}, \& {Bragaglia}}]{ses07}
{Sestito}, P., {Randich}, S., \& {Bragaglia}, A. 2007, \aap, 465, 185

\bibitem[{{Smolinski} {et~al.}(2011){Smolinski}, {Lee}, {Beers}, {An},
  {Bickerton}, {Johnson}, {Loomis}, {Rockosi}, {Sivarani}, \& {Yanny}}]{smo11}
{Smolinski}, J.~P., {et~al.} 2011, \aj, 141, 89

\bibitem[{{Spitoni} \& {Matteucci}(2011)}]{spi11}
{Spitoni}, E., \& {Matteucci}, F. 2011, \aap, 531, A72+

\bibitem[{{Stanghellini} \& {Haywood}(2010)}]{sta10}
{Stanghellini}, L., \& {Haywood}, M. 2010, \apj, 714, 1096

\bibitem[{{Steinmetz} {et~al.}(2006){Steinmetz}, {Zwitter}, {Siebert},
  {Watson}, {Freeman}, {Munari}, {Campbell}, {Williams}, {Seabroke}, {Wyse},
  {Parker}, {Bienaym{\'e}}, {Roeser}, {Gibson}, {Gilmore}, {Grebel}, {Helmi},
  {Navarro}, {Burton}, {Cass}, {Dawe}, {Fiegert}, {Hartley}, {Russell},
  {Saunders}, {Enke}, {Bailin}, {Binney}, {Bland-Hawthorn}, {Boeche}, {Dehnen},
  {Eisenstein}, {Evans}, {Fiorucci}, {Fulbright}, {Gerhard}, {Jauregi}, {Kelz},
  {Mijovi{\'c}}, {Minchev}, {Parmentier}, {Pe{\~n}arrubia}, {Quillen}, {Read},
  {Ruchti}, {Scholz}, {Siviero}, {Smith}, {Sordo}, {Veltz}, {Vidrih}, {von
  Berlepsch}, {Boyle}, \& {Schilbach}}]{ste06}
{Steinmetz}, M., {et~al.} 2006, \aj, 132, 1645

\bibitem[{{Stewart} {et~al.}(2008){Stewart}, {Bullock}, {Wechsler}, {Maller},
  \& {Zentner}}]{ste08}
{Stewart}, K.~R., {Bullock}, J.~S., {Wechsler}, R.~H., {Maller}, A.~H., \&
  {Zentner}, A.~R. 2008, \apj, 683, 597

\bibitem[{{Stinson} {et~al.}(2010){Stinson}, {Bailin}, {Couchman}, {Wadsley},
  {Shen}, {Nickerson}, {Brook}, \& {Quinn}}]{sti10}
{Stinson}, G.~S., {Bailin}, J., {Couchman}, H., {Wadsley}, J., {Shen}, S.,
  {Nickerson}, S., {Brook}, C., \& {Quinn}, T. 2010, \mnras, 408, 812

\bibitem[{{Twarog} {et~al.}(1997){Twarog}, {Ashman}, \&
  {Anthony-Twarog}}]{twa97}
{Twarog}, B.~A., {Ashman}, K.~M., \& {Anthony-Twarog}, B.~J. 1997, \aj, 114,
  2556

\bibitem[{{Venn} {et~al.}(2004){Venn}, {Irwin}, {Shetrone}, {Tout}, {Hill}, \&
  {Tolstoy}}]{ven04}
{Venn}, K.~A., {Irwin}, M., {Shetrone}, M.~D., {Tout}, C.~A., {Hill}, V., \&
  {Tolstoy}, E. 2004, \aj, 128, 1177

\bibitem[{{Villalobos} \& {Helmi}(2008)}]{vil08}
{Villalobos}, {\'A}., \& {Helmi}, A. 2008, \mnras, 391, 1806

\bibitem[{{Villanova} {et~al.}(2005){Villanova}, {Carraro}, {Bresolin}, \&
  {Patat}}]{vil05}
{Villanova}, S., {Carraro}, G., {Bresolin}, F., \& {Patat}, F. 2005, \aj, 130,
  652

\bibitem[{{Wilson} {et~al.}(2011){Wilson}, {Helmi}, {Morrison}, {Breddels},
  {Bienaym{\'e}}, {Binney}, {Bland-Hawthorn}, {Campbell}, {Freeman},
  {Fulbright}, {Gibson}, {Gilmore}, {Grebel}, {Munari}, {Navarro}, {Parker},
  {Reid}, {Seabroke}, {Siebert}, {Siviero}, {Steinmetz}, {Williams}, {Wyse}, \&
  {Zwitter}}]{wil11}
{Wilson}, M.~L., {et~al.} 2011, \mnras, 413, 2235

\bibitem[{{Yanny} {et~al.}(2009){Yanny}, {Rockosi}, {Newberg}, {Knapp},
  {Adelman-McCarthy}, {Alcorn}, {Allam}, {Allende Prieto}, {An}, {Anderson},
  {Anderson}, {Bailer-Jones}, {Bastian}, {Beers}, {Bell}, {Belokurov},
  {Bizyaev}, {Blythe}, {Bochanski}, {Boroski}, {Brinchmann}, {Brinkmann},
  {Brewington}, {Carey}, {Cudworth}, {Evans}, {Evans}, {Gates}, {G{\"a}nsicke},
  {Gillespie}, {Gilmore}, {Gomez-Moran}, {Grebel}, {Greenwell}, {Gunn},
  {Jordan}, {Jordan}, {Harding}, {Harris}, {Hendry}, {Holder}, {Ivans},
  {Ivezi{\v c}}, {Jester}, {Johnson}, {Kent}, {Kleinman}, {Kniazev},
  {Krzesinski}, {Kron}, {Kuropatkin}, {Lebedeva}, {Lee}, {Leger}, {L{\'e}pine},
  {Levine}, {Lin}, {Long}, {Loomis}, {Lupton}, {Malanushenko}, {Malanushenko},
  {Margon}, {Martinez-Delgado}, {McGehee}, {Monet}, {Morrison}, {Munn},
  {Neilsen}, {Nitta}, {Norris}, {Oravetz}, {Owen}, {Padmanabhan}, {Pan},
  {Peterson}, {Pier}, {Platson}, {Fiorentin}, {Richards}, {Rix}, {Schlegel},
  {Schneider}, {Schreiber}, {Schwope}, {Sibley}, {Simmons}, {Snedden}, {Smith},
  {Stark}, {Stauffer}, {Steinmetz}, {Stoughton}, {Subba Rao}, {Szalay},
  {Szkody}, {Thakar}, {Thirupathi}, {Tucker}, {Uomoto}, {Vanden Berk},
  {Vidrih}, {Wadadekar}, {Watters}, {Wilhelm}, {Wyse}, {Yarger}, \&
  {Zucker}}]{yan09}
{Yanny}, B., {et~al.} 2009, \aj, 137, 4377

\bibitem[{{Yoachim} \& {Dalcanton}(2005)}]{yoa05}
{Yoachim}, P., \& {Dalcanton}, J.~J. 2005, \apj, 624, 701

\bibitem[{{Yoachim} \& {Dalcanton}(2006)}]{yoa06}
---. 2006, \aj, 131, 226

\bibitem[{{Yoachim} \& {Dalcanton}(2008)}]{yoa08b}
---. 2008, \apj, 683, 707

\bibitem[{{Yong} {et~al.}(2005){Yong}, {Carney}, \& {Teixera de
  Almeida}}]{yon05}
{Yong}, D., {Carney}, B.~W., \& {Teixera de Almeida}, M.~L. 2005, \aj, 130, 597

\bibitem[{{Yong} {et~al.}(2006){Yong}, {Carney}, {Teixera de Almeida}, \&
  {Pohl}}]{yon06}
{Yong}, D., {Carney}, B.~W., {Teixera de Almeida}, M.~L., \& {Pohl}, B.~L.
  2006, \aj, 131, 2256

\bibitem[{{York} {et~al.}(2000){York}, {Adelman}, {Anderson}, {Anderson},
  {Annis}, {Bahcall}, {Bakken}, {Barkhouser}, {Bastian}, {Berman}, {Boroski},
  {Bracker}, {Briegel}, {Briggs}, {Brinkmann}, {Brunner}, {Burles}, {Carey},
  {Carr}, {Castander}, {Chen}, {Colestock}, {Connolly}, {Crocker}, {Csabai},
  {Czarapata}, {Davis}, {Doi}, {Dombeck}, {Eisenstein}, {Ellman}, {Elms},
  {Evans}, {Fan}, {Federwitz}, {Fiscelli}, {Friedman}, {Frieman}, {Fukugita},
  {Gillespie}, {Gunn}, {Gurbani}, {de Haas}, {Haldeman}, {Harris}, {Hayes},
  {Heckman}, {Hennessy}, {Hindsley}, {Holm}, {Holmgren}, {Huang}, {Hull},
  {Husby}, {Ichikawa}, {Ichikawa}, {Ivezi{\'c}}, {Kent}, {Kim}, {Kinney},
  {Klaene}, {Kleinman}, {Kleinman}, {Knapp}, {Korienek}, {Kron}, {Kunszt},
  {Lamb}, {Lee}, {Leger}, {Limmongkol}, {Lindenmeyer}, {Long}, {Loomis},
  {Loveday}, {Lucinio}, {Lupton}, {MacKinnon}, {Mannery}, {Mantsch}, {Margon},
  {McGehee}, {McKay}, {Meiksin}, {Merelli}, {Monet}, {Munn}, {Narayanan},
  {Nash}, {Neilsen}, {Neswold}, {Newberg}, {Nichol}, {Nicinski}, {Nonino},
  {Okada}, {Okamura}, {Ostriker}, {Owen}, {Pauls}, {Peoples}, {Peterson},
  {Petravick}, {Pier}, {Pope}, {Pordes}, {Prosapio}, {Rechenmacher}, {Quinn},
  {Richards}, {Richmond}, {Rivetta}, {Rockosi}, {Ruthmansdorfer}, {Sandford},
  {Schlegel}, {Schneider}, {Sekiguchi}, {Sergey}, {Shimasaku}, {Siegmund},
  {Smee}, {Smith}, {Snedden}, {Stone}, {Stoughton}, {Strauss}, {Stubbs},
  {SubbaRao}, {Szalay}, {Szapudi}, {Szokoly}, {Thakar}, {Tremonti}, {Tucker},
  {Uomoto}, {Vanden Berk}, {Vogeley}, {Waddell}, {Wang}, {Watanabe},
  {Weinberg}, {Yanny}, \& {Yasuda}}]{yor00}
{York}, D.~G., {et~al.} 2000, \aj, 120, 1579

\bibitem[{{Yoshii}(1982)}]{yos82}
{Yoshii}, Y. 1982, \pasj, 34, 365

\end{thebibliography}

\end{document}